\documentclass[11pt]{article}
\usepackage{amsmath,amssymb,color,epsfig,cite,bm}
\usepackage{graphicx}
\usepackage{subfigure}
\usepackage{setspace}
\usepackage{amsfonts}
\newcommand{\hoch}[1]{$\, ^{#1}$}

\makeatletter
\@addtoreset{equation}{section}
\makeatother

\newcommand{\be}{\begin{equation}}
\newcommand{\ee}{\end{equation}}
\newcommand{\bea}{\setlength\arraycolsep{2pt} \begin{eqnarray}}
\newcommand{\eea}{\end{eqnarray}}
\newcommand{\nn}{\nonumber}

\def\ft#1#2{{\textstyle{\frac{\scriptstyle #1}{\scriptstyle #2} } }}
\def\fft#1#2{{\frac{#1}{#2}}}
\def\CP{{{\mathbb C}{\mathbb P}}}
\def\0{{\sst{(0)}}}
\def\1{{\sst{(1)}}}
\def\2{{\sst{(2)}}}
\def\3{{\sst{(3)}}}
\def\4{{\sst{(4)}}}
\def\5{{\sst{(5)}}}
\def\6{{\sst{(6)}}}
\def\7{{\sst{(7)}}}
\def\8{{\sst{(8)}}}
\def\sst#1{{\scriptscriptstyle #1}}

\def\del{{\partial}}

\def\AA{{{\sst A}}}

\def\im{{{\rm i\,}}}
\def\R{{\mathbb R}}

\def\bx{{\bf x}}
\def\bd{{\bf d}}

\begin{document}

\begin{flushright}
\hfill{MI-TH-1885}

\end{flushright}

\begin{center}
{\Large {\bf Super-Geometrodynamics in Higher Dimensions}}

\vspace{15pt}
{\Large Surya Kiran Kanumilli\hoch{1} and C.N. Pope\hoch{1,2}}

\vspace{15pt}

\hoch{1}{\it George P. \& Cynthia Woods Mitchell  Institute
for Fundamental Physics and Astronomy,\\
Texas A\&M University, College Station, TX 77843, USA}

\vspace{10pt}

\hoch{2}{\it DAMTP, Centre for Mathematical Sciences,
 Cambridge University,\\  Wilberforce Road, Cambridge CB3 OWA, UK}

\vspace{30pt}

\underline{ABSTRACT}

\end{center}

   The geometrodynamics of the four-dimensional Einstein and 
Einstein-Maxwell theories were first studied by Wheeler and Misner more
than fifty years ago, by constructing solutions of the constraints on
an initial spatial slice in a Hamiltonian formulation of the theories.
More recently these considerations were extended to various 
four-dimensional theories with additional fields, encompassing cases
that arise in supergravity and the low-energy limit of compactified 
string theory.  In this paper we extend these considerations further, by
constructing solutions of the initial value constraints in higher spacetime
dimensions, for wide classes of theories that include supergravities and the
low-energy limits of string and M-theory.  We obtain time-symmetric 
initial data sets for multiple black hole spacetimes and also wormholes
in higher dimensions.

\vfill {\footnotesize Emails: 
suryak@physics.tamu.edu\ \ \ pope@physics.tamu.edu}

\thispagestyle{empty}

\pagebreak

\tableofcontents
\addtocontents{toc}{\protect\setcounter{tocdepth}{2}}

%%%%%%%%%%%%%%%%%%%%%%%%%%%%%%%%%%%%%%%%

%\newpage
%%%%%%%%%%%%%%%%%%%%%%%%%%%%%%%%%%%%%%%%

\section{Introduction}

   The initial value formulation of general relativity provides a powerful
tool for studying time-dependent solutions, especially in situations where one
cannot solve the equations explicitly.  The initial value constraints can place
restrictions on the possible topologies and geometries of the Cauchy surface
on which the initial conditions for the problem are defined.  The approach
was largely pioneered by Wheeler \cite{Wheeler} and Misner 
\cite{MisnerWheeler,Misner:1960zz,Misner2}, under the name of 
{\it Geometrodynamics}, with further early developments
by Lindquist, and Brill \cite{Lindquist,Brill:1963yv} and other authors.

   The initial value formulation has subsequently been extended to larger
systems of matter coupled to Einstein gravity, and among these, the theories
encountered in supergravities are of particular interest.  First of all,
such extensions of Einstein gravity 
have the merit of being consistent with the usual positive
energy theorems of classical general relativity.  Furthermore, they are
potentially of intrinsic physical interest if they arise as low energy 
limits of string theory or
M-theory.  Some results in one such theory, namely a particular 
four-dimensional Einstein-Maxwell-Dilaton
(EMD) theory coming from string theory, were studied by Ortin \cite{ortin}.  
In \cite{cvegibpop} a larger class of EMD theories and other string-related 
supergravities were investigated, with an emphasis on time-symmetric 
initial data sets.  The focus in \cite{cvegibpop} was exclusively on
four-dimensional theories.  Clearly, in the context of supergravity, 
string theory and M-theory it is of interest to extend the investigation to
dimensions greater than four, and that provides the motivation for the
present paper.  As in \cite{cvegibpop} we shall, for simplicity, focus on
the case of time-symmetric initial data; i.e., on the case where the the
the metric is taken to be static on the initial time surface, and the
second fundamental form vanishes there.

  The bulk of the earlier studies in four dimensions involved making an
ansatz introduced by Lichnerowicz \cite{lichnerowicz}, in which the
spatial metric on the initial surface was taken to be a conformal factor
times a fiducial static metric $\bar g_{ij}$, where $\bar g_{ij}$ might
typically be the flat Euclidean metric, or the metric on the 3-sphere, or
the metric on $S^1\times S^2$.  The Hamiltonian constraint now becomes 
an equation for the spatial Laplacian of the conformal factor, allowing
rather simple solutions if a suitable restriction of its coordinate 
dependence, adapted to the symmetry of the fiducial metric, is imposed. By this
means data sets that describe the initial data for multiple black holes,
or black holes in a closed universe, or wormholes in the $S^1\times S^2$ case, 
can be constructed.

  In higher dimensions the possibilities for choosing fiducial metrics
$\bar g_{ij}$ become more extended.  For example, if there are $d$
spatial dimensions one can write the metric on a round $d$-sphere $S^d$
in a variety of different ways, such as in terms of foliations of 
$S^p\times S^q$ surfaces with a ``latitude'' coordinate $\mu$, where
$p+q=d-1$.  If one then makes an assumption that the conformal factor
depends only on $\mu$, then depending upon how the integers $p$ and $q$ 
partition $d-1$, one will obtain solutions of the initial data 
constraints corresponding to different distributions of black hole centres.
There are also many possibilities extending the $S^1\times S^2$ wormhole
choice that was considered when $d=3$.

  In this paper, we shall consider some of these higher-dimensional
generalisatiions in some detail.  After setting up the notation for
time-symmetric initial data in section 2, we turn in section 3 to the 
case of the higher-dimensional vacuum Einstein equations.  We
construct solutions of the initial-value constraints both for a flat
Euclidean fiducial metric $\bar g_{ij}$, and for cases with a spherical
metric, described in a variety of different ways as described above.  
We also consider one example, for $d=4$, where the fiducial metric is 
taken to be 
the Fubini-Study metric on the complex projective plane $\CP^2$.   
In section 4 we consider the Einstein-Maxwell equations, extending this to
the Einstein-Maxwell-Dilaton system in section 5.  Another generalisation,
to an Einstein-Dilaton system coupled to two electromagnetic fields, is
considered in section 6.  In section 7 we give a rather general discussion
of Einstein gravity coupled to $p$ dilatons and $q$ Maxwell fields.  This
encompasses many examples that arise in supergravities in various dimensions.
We describe in section 8 how the initial-value problem for these
general Einstein-Maxwell-Dilaton systems may be mapped into the initial-value
problem for corresponding purely Einstein-Dilaton systems.

  In section 9, we turn to a consideration of initial data for wormhole 
solutions in higher dimensions, generalising results in the literature on
the $d=3$ case.  We consider wormholes associated with using
a fiducial metric on $S^1\times S^{d-1}$.  We obtain solutions for 
wormhole initial data in higher-dimensional pure Einstein, 
Einstein-Maxwell, and the various Einstein-Maxwell-Dilatons systems 
mentioned above.  We include in our discussion a calculation of
the masses and the charges for these wormhole configurations, and 
the interaction energies between multiple wormhole
throats.  The paper ends with conclusions in section 10.

\section{The Constraints for Time-Symmetric Initial Data}

  In the general ADM decomposition, an $n$-dimensional metric $d\hat s^2$
is written in the form
%%%%%
\be
d\hat s^2= -N^2\, dt^2 + g_{ij}\, (dx^i + N^i\, dt)(dx^j + N^j\, dt)\,,
\ee
%%%%%
whose inverse is given by
%%%%%
\be
\Big(\fft{\del}{\del \hat s}\Big)^2 = -\fft{1}{N^2}\, 
\Big(\fft{\del}{\del t} - N^i \del_i\Big)^2 + g^{ij}\, \del_i\otimes \del_j
\,.
\ee
%%%%%
The unit vector normal to the $t=\,$constant surfaces is given by
%%%%%
\be
n= n^\mu\, \del_\mu = \fft1{N}\, \Big(\fft{\del}{\del t} - N^i \del_i\Big)\,.
\ee
%%%%%

    The Hamiltonian constraint for the $n$-dimensional 
Einstein equations $\hat R_{\mu\nu}- \ft12 \hat R\, \hat g_{\mu\nu} 
= \hat T_{\mu\nu}$ is
then given by the double projection with $n^\mu\, n^\nu$:
%%%%%
\be
\hat R_{\mu\nu}\, n^\mu\, n^\nu + \ft12 \hat R = 
  \hat T_{\mu\nu}\, n^\mu\, n^\nu\,.
\ee
%%%%%
By the Gauss-Codacci equations this implies
%%%%%
\be
R + K^2 - K^{ij}\, K_{ij} = 2 \hat T_{\mu\nu}\, n^\mu\, n^\nu\,,
\ee
%%%%%
where $R$ is the Ricci scalar of the $(n-1)$-dimensional spatial metric
$g_{ij}$ and $K_{ij}$ is the second fundamental form.
If we consider time-symmetric data on the initial surface, which we take
to be at $t=0$, then
$K_{ij}=0$ and $N^i=0$ on this surface, and the Hamiltonian constraint becomes
%%%%%
\be
R = 2 N^{-2}\, \hat T_{00}\,.\label{Hamcon}
\ee
%%%%%
Note that the momentum constraint will simply be $\hat T_{0i}=0$.

   Following a procedure introduced by Lichnerowicz \cite{lichnerowicz},
we may seek solutions to the Hamiltonian constraint by considering the
case where the metric $g_{ij}$ is conformally related to a fixed,
time-independent background metric $\bar g_{ij}$, with $g_{ij}=
\Phi^\alpha\, \bar g_{ij}$.  It is straightforward to see that if
we choose $\alpha= 4/(d-2)$ then we shall have
%%%%%
\be
R = \Phi^{-\, \fft{d+2}{d-2}}\, \Big[ -\fft{4(d-1)}{d-2}\, \bar\square +
   \bar R\Big]\, \Phi\,,\qquad 
  g_{ij}= \Phi^{\fft{4}{d-2}}\, \bar g_{ij}\,,\label{lichR}
\ee
%%%%%
where $\bar R$ is the Ricci scalar of the background metric $\bar g_{ij}$,
and $\bar\square$ is the covariant Laplacian $\bar\nabla^i\bar\nabla_i$ 
in the background metric.  Here, and in what follows, we are using 
$d$ to denote the number of spatial dimensions, so
%%%%%
\be
  d=n-1\,.
\ee
%%%%%

\section{Data for Vacuum Einstein Equations}

   If we consider the constraint equations for the pure vacuum Einstein 
equations then we shall simply have the Hamiltonian constraint $R=0$ on
the initial $t=0$ surface which, from (\ref{lichR}), will give the linear
equation
%%%%%
\be
-\bar\square\Phi + \fft{d-2}{4(d-1)}\,\bar R\,  \Phi=0\label{Philin}
\ee
%%%%%
for $\Phi$.  Any solution of this equation will give rise to consistent 
time-symmetric initial data for the vacuum Einstein equations.  Because
the equation is linear in $\Phi$, one can of course superpose solutions.

   In the bulk of this paper where we consider various matter couplings
to gravity we shall study the simplest case
where the background metric is just taken to be flat, with $\bar g_{ij}=
\delta_{ij}$.  Before doing so, in this section we shall also make some 
observations
about the vacuum Einstein case with more complicated curved background metrics.

\subsection{Vacuum data with flat $\bar g_{ij}$}\label{vacuumflatsec}

  The simplest choice for the background metric $\bar g_{ij}$ in (\ref{lichR})
is to take it to be flat, with $\bar g_{ij}=\delta_{ij}$.  We then get
vacuum initial data by taking $\Phi$ to be any harmonic function in the
flat metric, obeying 
%%%%%
\be
\del_i\del_i\, \Phi=0\,.
\ee
%%%%%
We may therefore take $\Phi$ to be of the form
%%%%%
\be
\Phi = 1 + \ft12 \sum_{n=1}^N \fft{M_n}{|\bx - \bx_n|^{d-2}}\,,
\ee
%%%%%
where $\bx$ denotes the $(d-1)$-vector $\bx=(x^1,x^2,\cdots ,x^{d-1})$.  

   In general, the case with $N$ centres corresponds to initial data for
a system of $N$ black holes, which would, of course, evolve as a time-dependent
solution, which one could solve numerically but solving explicitly would
not be tractable.  The $N=1$ case with a single singularity, however, 
simply gives the initial data for the
$(d+1)$-dimensional Schwarzschild solution.  Taking the singularity,
without loss of generality, to be at the origin (so $\bx_1=0$), and
taking $M_1=M$, then in terms of hyperspherical polar coordinates 
in the Euclidean $d$-space we have
%%%%%
\bea
g_{ij}\, dx^i dx^j &=& \Phi^{\fft4{d-2}}\, \Big(d\rho^2 + \rho^2\, 
   d\Omega_{d-1}^2\Big)\,,\nn\\
\Phi &=& 1 +\fft{M}{2 \rho^{d-2}}\,,\label{schwid}
\eea
%%%%%
where $\rho^2 = x^i x^i$ and $d\Omega_{d-1}^2$ is the metric on the unit
$(d-1)$-sphere.   To see how this corresponds to the initial data for
the Schwarzschild solution, we observe that the metric in (\ref{schwid}) 
can be written in the standard $d$-dimensional Schwarzschild form
%%%%%
\be
g_{ij}\, dx^i dx^j = \Big(1-\fft{2M}{r^{d-2}}\Big)^{-1}\, dr^2 +
   r^2\, d\Omega_{d-1}^2
\ee
%%%%%
if 
%%%%%
\be
r= \rho\, \Big(1+ \fft{M}{2\rho^{d-2}}\Big)^{\fft{2}{d-2}}\quad \hbox{and}\quad
\Big(1-\fft{2M}{r^{d-2}}\Big)^{-1}\, dr^2= 
\Big(1 +\fft{M}{2 \rho^{d-2}}\Big)^{\fft{4}{d-2}}\, d\rho^2\,.\label{rrhorel}
\ee
%%%%%
A straightforward calculation shows that indeed if $r$ is given in terms of
$\rho$ by the first equation in (\ref{rrhorel}), then the second equation is
satisfied too.  Note that the first equation can be inverted to give 
$\rho$ as a function of $r$, with the result
%%%%%
\be
\rho = r\, 
 \Big[\ft12 \Big(1 + \sqrt{1-\fft{2M}{r^{d-2}}}\Big)\Big]^{\fft2{d-2}}\,.
\ee
%%%%%

\subsection{Vacuum data with non-flat background $\bar g_{ij}$}

   In the case of four spacetime dimensions, the Lichnerowicz procedure
has been used in a variety of applications, with the background metric 
$\bar g_{ij}$ being taken to be either flat, or else the standard metric 
on the unit 3-sphere or on $S^1\times S^2$.  These latter cases have been 
used to construct initial data for black holes in closed universes, or for
wormholes.  In higher dimensions the possibilities for the choice of the
background metric $\bar g_{ij}$ are more diverse, and there are many cases
where one can solve explicitly for $\Phi$ on the initial $t=0$ surface.
(Of course, it does not necessarily mean that the data will evolve into
desirable solutions, but it does provide interesting cases for further
investigation.)  In what follows, we present some simple examples of
curved background metrics. 

\subsubsection{Unit $d$-sphere background}

  To illustrate some of the possibilities in higher dimensions, let us first
consider the case when the background metric $\bar g_{ij}$ describes the
unit $d$-sphere.  There are many ways that this background metric can be
written; here, we shall consider the cases
%%%%%
\be
\bar g_{ij}\, dx^i\, dx^j = d\bar s^2 = d\mu^2 + \sin^2\mu\, d\Omega_p^2 +
   \cos^2\mu\, d\widetilde \Omega_q^2\,,\qquad p+q=d-1\,,\label{dsphere}
\ee
%%%%%
where $d\Omega_p^2$ and $d\widetilde \Omega_q^2$ are unit metrics on 
a $p$-sphere and $q$-sphere respectively.  The ``latitude'' coordinate
$\mu$ ranges from $0\le\mu\le\ft12\pi$, except when $q=0$ when it ranges
over $0\le\mu\le \pi$, and $p=0$ when it ranges over 
$-\ft12\pi\le\mu\le\ft12\pi$. 

    The Ricci scalar of the unit $d$-sphere is given
by $\bar R= d\, (d-1) $, and so the
equation (\ref{Philin}) for $\Phi$ is the Helmholtz equation\footnote{Note
that the hyperspherical harmonics on the unit $d$-sphere obey
$-\bar\square Y =\lambda Y$ with eigenvalues $\lambda=\ell\,(\ell+d-1)$
and  $\ell=0,1,2,\ldots$, and since none of these eigenvalues coincide
with the eigenvalue in (\ref{helmholtz}) (which is in fact negative), 
 the solutions for $\Phi$ that we are
seeking will necessarily have singularities on the sphere.}
%%%%%
\be
-\bar\square\Phi + \ft14 d\, (d-2)\, \Phi=0\,.\label{helmholtz}
\ee
%%%%%
A simple ansatz for solving this explicitly in the (\ref{dsphere}) metrics
is to assume $\Phi$ is a function only of the latitude coordinate $\mu$, and
so (\ref{helmholtz}) becomes
%%%%%
\be
\Phi'' + (p\, \cot\mu - q\, \sin\mu)\, \Phi' -\ft14(p+q+1)(p+q-1)\, \Phi=0\,,
\label{pqeqn}
\ee
%%%%%
where a prime denotes a derivative with respect to $\mu$.  

   If we consider the simplest case where $q=0$ and hence $d=p+1$, the unit
$d$-sphere is viewed as a foliation by $(d-1)$-spheres.  The solution
to (\ref{pqeqn}) can be written as
%%%%
\be
\Phi = \fft{c_1}{(\cos\ft12\mu)^{p-1}} + \fft{c_2}{(\sin\ft12\mu)^{p-1}}\,,
\ee
%%%%%
where $c_1$ and $c_2$ are arbitrary constants. The first term has a 
singularity at the north pole, and the second term is singular at the
south pole.  If we choose $c_1=c_2= \sqrt{M}\, 2^{-\ft12 (p-1)}$, so that
%%%%%
\be
\Phi= \sqrt{\fft{M}{2}}\, 2^{-\ft12 (p-1)}\, \Big[\fft1{(\cos\ft12\mu)^{p-1}} +
     \fft1{(\sin\ft12\mu)^{p-1}}\Big]\,,\label{schwid2}
\ee
%%%%%
then after defining a new radial variable $r$ by letting
%%%%%
\be
r^{\ft12 (p-1)} = \sqrt{\fft{M}{2}}\,  \Big[ (\tan\ft12\mu)^{\ft12(p-1)} 
  + (\cot\ft12\mu)^{\ft12(p-1)}\Big]\,,
\ee
%%%%%
the spatial $d$-metric, given as in (\ref{lichR}), becomes
%%%%%
\be
ds^2= \Phi^{\ft{4}{d-2}}\, (d\mu^2 + \sin^2\mu\, d\Omega_{d-1}^2) =
   \Big( 1- \fft{2M}{r^{d-2}}\Big)^{-1}\, dr^2 + 
   r^2\, d\Omega_{d-1}^2\,.
\ee
%%%%%
This can be recognised as the time-symmetric initial data for the
$(d+1)$-dimensional generalisation of the Schwarzschild black hole.
The horizon of the black hole corresponds to the equator, $\mu=\ft12\pi$. 

  Of course since the metric on a round $d$-sphere is conformally related
to the flat Euclidean $d$-metric in any dimension, we can straightforwardly 
relate the
initial data we constructed here to the previous initial data for the
Schwarzschild black hole that we constructed in section \ref{vacuumflatsec}
using a flat background metric.  The Euclidean and sphere metrics are
related by
%%%%%
\be
d\rho^2 + \rho^2\, d\Omega_{d-1}^2 = \Omega^2\, \big( d\mu^2 +
  \sin^2\mu\, d\Omega_{d-1}^2\big)
\ee
%%%%%
where
%%%%%
\be
\Omega= \fft{c^2}{\cos^2\ft12\mu}\,, \qquad \rho = 2 c^2 \tan\ft12\mu\,,
\ee
%%%%%
where $c$ is an arbitrary constant, 
and using this one can easily verify that the $\Phi$ functions given in
(\ref{schwid}) and (\ref{schwid2}) for the flat and the spherical 
background metrics are related by
%%%%%
\be
 \Phi_{\rm sphere} = \Phi_{\rm flat}\, \Omega^{\fft{d-2}{2}}\,.
\ee
%%%%%

  In the manner described, for example, in \cite{LW,clifton}, one could take 
a superposition of solutions for $\Phi$ like the one in (\ref{schwid2}), 
but expressed
in terms of a rotated choice of polar axes for the $d$-sphere.  By this means,
one could set time-symmetric initial data for multiple black holes.

   We can also consider the solutions of the equation (\ref{pqeqn}) for $\Phi$ 
in the case where $p$ and $q$ are both non-zero.  In this description of
the $d=p+q+1$ sphere it is foliated by $S^p\times S^q$ surfaces. The
solutions are given in terms of hypergeometric functions by
%%%%%
\bea
\Phi &=& 
c_1 \, F\Big[\fft{p+q-1}{4},\fft{p+q+1}{4},\fft{q+1}{2}; \cos^2\mu\Big]
\nn\\
&&
+ c_2\, (\cos\mu)^{1-q}\, F\Big[\fft{p-q+1}{4},\fft{p-q+3}{4}, \fft{3-q}{2};
  \cos^2\mu\Big]\,.
\eea
%%%%% 
The explicit solutions are of varying complexity depending on the choice of
$p$ and $q$.  A fairly simple example is when $p=q=2$.  In this case,
the solution to (\ref{pqeqn}) for $\Phi$ for this metric on the unit
5-sphere is given by
%%%%%
\be
\Phi= \fft1{\cos\mu}\, \Big(\fft{a_1}{\sin\ft12\mu} + \fft{a_2}{\cos\ft12\mu}
\Big)\,.\label{bhid}
\ee
%%%%%
This has a simple pole singularity in a 3-plane times a 2-sphere surface 
at $\mu=0$ and a simple pole singularity on another 3-plane times 2-sphere
surface at $\mu=\ft12\pi$.  (Recall here the 5-sphere is spanned by 
$0\le\mu\le\ft12\pi$, with a foliation of different 2-spheres
contracting onto the origin of a 3-plane at each endpoint.)  The
initial data described by (\ref{bhid}) would correspond to taking 
certain continuous superpositions of elementary mass-point initial data of
the kind we discussed previously.  

\subsubsection{$\CP^2$ background}

   Further possibilities for background metrics that could give 
explicitly solvable time-symmetric initial data include taking $\bar g_{ij}$
to be a metric on a product of spheres, or else taking a metric on
a complex projective space or products of these, possibly with spheres as
well.  As a simple example, consider the case $d=4$ with $\bar g_{ij}$
taken to be the Fubini-Study metric on $\CP^2$.  The metric
%%%%%
\be
d\bar s^2= d\mu^2 + \ft14 \sin^2\mu\, \cos^2\mu\, (d\psi+\cos\theta\, d\phi)^2
+ \ft14 \sin^2\mu\, (d\theta^2 + \sin^2\theta\, d\phi^2)\label{cp2met}
\ee
%%%%%
is Einstein with $\bar R_{ij}= 6 \bar g_{ij}$ and hence $\bar R=24$. 
 From (\ref{Philin}) we have $-\bar\square\Phi + 4\Phi=0$, and so if we assume
$\Phi$ depends only on $\mu$ we have
%%%%%
\be
\Phi'' + (3\cot\mu-\tan\mu)\, \Phi' - 4\Phi=0\,,
\ee
%%%%%
for which the general solution is
%%%%%
\be
\Phi= \fft{c_1}{\sin^2\mu} + \fft{c_2\, \log\cos\mu}{\sin^2\mu}\,.
\label{cp2Phi}
\ee
%%%%%
The first term exhibits the leading-order $1/\mu^2$ behaviour of a point 
charge at the NUT at $\mu=0$, while the second term has the characteristic
$\log(\ft12\pi-\mu)$ behaviour of a charge distributed over the bolt at 
$\mu=\ft12\pi$.  It would be interesting to investigate what this 
time-symmetric initial data describes in this, and other cases.

\section{Einstein-Maxwell Equations}\label{EMsec}

   Consider the Einstein-Maxwell system in $n=(d+1)$ dimensions, described
by the action
%%%%%
\be
I= \int d^nx \sqrt{-\hat g} (\hat R - \hat F^2)\,,
\ee
%%%%%
where $\hat F^2= \hat g^{\mu\rho}\,\hat g^{\nu\sigma}\, 
\hat F_{\mu\nu}\, \hat F_{\rho\sigma}$, 
for which the equations of motion are
%%%%%
\be
\hat R_{\mu\nu} -\ft12\hat R\, \hat g_{\mu\nu} = 2(\hat F_{\mu\rho}\, 
  \hat F_{\nu\sigma} \hat g^{\rho\sigma} -\ft14 \hat F^2\, \hat g_{\mu\nu})\,,
\qquad \hat \nabla_\mu \hat F^{\mu\nu}=0\,.\label{einstmaxeqn}
\ee
%%%%%
We shall consider the evolution of purely electric time-symmetric
initial data, for which only the components of $\hat F_{\mu\nu}$ specified
by
%%%%%
\be
n^\mu \, \hat F_{\mu i}= -E_i
\ee
%%%%%
are non-zero.  Projecting the Einstein equations in (\ref{einstmaxeqn}) with
$n^\mu n^\nu$ then gives the Hamiltonian constraint
%%%%%
\be
  R= 2 E^2\,,\label{Hamconeinmax}
\ee
%%%%%
where $E^2 = g^{ij}\, E_i\, E_j$.  We also have the Gauss law constraint
$\nabla_i E^i=0$.  

   Generalising a discussion of Misner and
Wheeler \cite{MisnerWheeler} to the case of general dimensions, we may seek 
a solution of the constraint equations, with $g_{ij}$ given as in 
(\ref{lichR}) and $\bar g_{ij}=\delta_{ij}$ as before, in the form
%%%%%
\be
\Phi= (CD)^\alpha\,,\qquad E_i= \beta\, \del_i\log\fft{C}{D}\,.\label{PhiE}
\ee
%%%%%
The aim is to choose the constants $\alpha$ and $\beta$ appropriately, such
that the constraint equations will be satisfied if $C$ and $D$ are arbitrary
harmonic functions in the flat background Euclidean metric $\delta_{ij}$.
Substituting $R$ from (\ref{lichR}) (with $\bar R=0$ since we are taking a
flat background metric $\bar g_{ij}=\delta_{ij}$) into (\ref{Hamconeinmax}),
together with (\ref{PhiE}), it is straightforward to see that if we
choose
%%%%%
\be
\alpha = \fft12\,,\qquad \beta = \sqrt{\fft{d-1}{2(d-2)}}\,,
\ee
%%%%%
then the Hamiltonian constraint is indeed satisfied if $C$ and $D$ are 
arbitrary harmonic functions in the Euclidean background metric $\bar g_{ij}=
\delta_{ij}$, i.e. 

%%%%%
\be
\del_i\del_i C=0\,,\qquad \del_i\del_i D=0\,.\label{CDharm}
\ee
%%%%%
Furthermore, the Gauss law constraint $\nabla_i E^i=0$ is also satisfied 
subject to (\ref{CDharm}).  

   If we take the case where $C$ and $D$ both have a single pole at the
origin, then in hyperspherical polar coordinates we can take
%%%%%
\be
C= 1 +\fft{M-Q}{2\rho^{d-2}}\,,\qquad D= 1 + \fft{M+Q}{2\rho^{d-2}}\,.
\ee
%%%%%
The spatial metric takes the form
%%%%%
\be
g_{ij} dx^i dx^j = (C D)^{\fft{2}{d-2}}\, (d\rho^2 + \rho^2 d\Omega_{d-1}^2)\,,
\ee
%%%%%
and if we define the area coordinate $r=\rho\, (CD)^{1/(d-2)}$ this becomes
%%%%%
\be
g_{ij} dx^i dx^j = \Big(1- \fft{2M}{r^{d-2}} + \fft{Q^2}{r^{2d-4}}\Big)^{-1}\,
  dr^2 + r^2 d\Omega_{d-1}^2\,,
\ee
%%%%%
which can be recognised as the spatial part of the $(d+1)$-dimensional
Reissner-Nordstr\"om metric.

\section{Einstein-Maxwell-Dilaton System}\label{EMDsec}

  A rather general class of theories that are relevant in string theory
are encompassed by the Einstein-Maxwell-Dilaton (EMD) system in $n=d+1$ 
dimensions, described by the Lagrangian
%%%%%
\be
{\cal L}= \sqrt{-\hat g}\, (\hat R -\ft12 \hat g^{\mu\nu}\, \del_\mu\phi
\del_\nu\phi - \ft14 e^{a\phi}\, \hat F^2)\,,
\ee
%%%%%
where $a$ is an arbitrary constant.  
The Hamiltonian and Gauss law constraints will be
%%%%%
\bea
   R &=&\ft12 g^{ij}\del_i\phi\, \del_j\phi + \ft12 e^{a\phi}\, g^{ij}\,
         E_i E_j\,,\nn\\
0&=& \nabla_i(e^{a\phi}\, g^{ij}\, E_j)\,.\label{emdcons}
\eea
%%%%%
As usual, we shall consider a flat Euclidean background metric, with
$g_{ij}= \Phi^{4/(d-2)}\, \delta_{ij}$.  
Upon using (\ref{lichR}) the Hamiltonian constraint becomes
%%%%%
\be
\fft{4(d-1)}{(d-2)}\, \del_i\del_i \Phi +
  \ft12 \Phi\, \Big[ \del_i\phi\, \del_i\phi + e^{a\phi}\, E_i E_i\Big]=0\,.
\ee
%%%%%

  It is straightforward to verify that we can solve the Hamiltonian and Gauss
law constraints by introducing three arbitrary harmonic functions 
$C$, $D$ and $W$ in the Euclidean $d$-space, in terms of which we write
%%%%%
\bea
  \Phi &=& (CD)^{\fft{(d-2)}{(d-1)\, \Delta}}\,W^{\fft{a^2}{\Delta}}\,,\nn\\
e^{a\phi} &=& \Big(\fft{CD}{W^2}\Big)^{\fft{2a^2}{\Delta}}\,,\nn\\
E_i &=& \fft{2}{\sqrt\Delta}\, \Big(\fft{CD}{W^2}\Big)^{-\fft{a^2}{\Delta}}\,
  \del_i\log\fft{C}{D}\,,\label{emddata}
\eea
%%%%%
where we have introduced the parameter $\Delta$, as in \cite{lupopemaximal},
which is related to $a$ by the expression
%%%%%
\be
a^2 = \Delta - \fft{2(d-2)}{d-1}\,.\label{Deltaparam}
\ee
%%%%%

\subsection{Non-extremal black hole}

  The solution for a static non-extremal black hole in the EMD theory
with arbitrary dilaton coupling $a$ in $n=d+1$ dimensions 
was constructed in \cite{emdbh}, and is given by
%%%%%
\bea
ds^2 &=& - h \, f^{-2(d-2)}\, dt^2 + f^2\, h^{-1}\, dr^2 +
   r^2\, f^2\, d\Omega_{d-1}^2\,,\nn\\
h&=& 1 - \Big(\fft{r_H}{r}\Big)^{d-2}\,,\qquad
  f= \Big(1 + \fft{\alpha}{r^{d-2}}\Big)^{\ft{2}{(d-1)\Delta}}\,,\nn\\
e^{a\phi} &=& \Big(1 + \fft{\alpha}{r^{d-2}}\Big)^{\ft{2 a^2}{\Delta}}\,,
\qquad
A = \fft2{\sqrt\Delta}\, \sqrt{1+\fft{r_H^{d-2}}{\alpha}}\,
\Big(1+ \fft{\alpha}{r^{d-2}}\Big)^{-1}\, dt
\,,\label{emdbh}
\eea
%%%%%
where $\Delta$ is defined in (\ref{Deltaparam}).

   Defining a new radial coordinate $\rho$ by
%%%%%
\be
r^{d-2}= \rho^{d-2}\, \Big(1+\fft{uv}{\rho^{d-2}}\Big)^2\,,
\ee
%%%%%
where $u$ and $v$ are constants related to the horizon radius $r_H$ and
the parameter $\alpha$ by
%%%%%
\be
r_H^{d-2}= 4 u v\,,\qquad \alpha = (u-v)^2\,,
\ee
%%%%%
a straightforward calculation show that the metric (\ref{emdbh}) becomes
%%%%%
\be
ds^2 = -N^2\, dt^2 + \Phi^{\ft{4}{d-2}}\, 
             (d\rho^2 + \rho^2\, d\Omega_{d-1}^2)\,,\label{staticmet}
\ee
%%%%%
with 
%%%%%
\bea
\Phi &=& \Big[ \Big(1 + \fft{u^2}{\rho^{d-2}}\Big)
            \Big(1 + \fft{v^2}{\rho^{d-2}}\Big)\Big]^{\ft{d-2}{(d-1)\Delta}}
\, \Big(1 + \fft{uv}{\rho^{d-2}}\Big)^{\ft{a^2}{\Delta}}\,,\nn\\
N &=& \Big(1 - \fft{uv}{\rho^{d-2}}\Big)^2\,
 \Big[ \Big(1 + \fft{u^2}{\rho^{d-2}}\Big)
            \Big(1 + \fft{v^2}{\rho^{d-2}}\Big)
             \Big]^{-\fft{2(d-2)}{(d-1)\Delta}}\,
  \Big(1 + \fft{uv}{\rho^{d-2}}\Big)^{\ft{4(d-2)}{(d-1)\Delta} -1}\,.
\eea
%%%%%
Comparing with (\ref{emddata}), we see that the static non-extremal
black hole is generated by starting from the initial data in which the
harmonic functions $C$, $D$ and $W$ are taken to be
%%%%%
\be
C=1 +\fft{u^2}{\rho^{d-2}}\,,\qquad
D=1 +\fft{v^2}{\rho^{d-2}}\,,\qquad
W=1 +\fft{uv}{\rho^{d-2}}\,.
\ee
%%%%%

   If one takes more general solutions for the intial data, with $C$,
$D$ and $W$ having singularities at multiple locations, the
evolution would give rise to time-dependent solutions that could be
constructed only numerically.    However, if one takes very specific
initial data with multiple singularities, it can give rise to static 
solutions.  This will happen in the case of initial data for the 
multi-centre extremal nlack holes, discussed below:

\subsection{Multi-centre extremal black holes}

  The static multi-centre extremal black holes in $n=d+1$ dimensions 
are given by
%%%%%
\bea
ds^2 &=& - C^{\ft{4(d-2)}{(d-1)\Delta}}\, dt^2 + C^{\ft{4}{(d-1)\Delta}}\,
dy^i\, dy^i\,,\nn\\
A&=& \fft{2}{\sqrt{\Delta}}\, C^{-1}\, dt\,,\qquad
  e^{a\phi} = C^{\ft{2 a^2}{\Delta}}\,,
\eea
%%%%%
where $C$ is an arbitrary harmonic function in the $d$-dimensional
Euclidean space with metric $dy^i\, dy^i$.  Comparison with (\ref{emddata})
shows that indeed the harmonic function $C$ provides the initial data
for these solutions, with $D=W=1$.

\section{Einstein-Two-Maxwell-Dilaton System}\label{E2MDsec}

   An extension of the Einstein-Maxwell-Dilaton system containing two 
Maxwell fields, with just one dilaton, is of considerable interest.
The theory, which we shall refer to by the acronym E2MD, has the
Lagrangian
%%%%%
\be
{\cal L}= \sqrt{-g}\, \Big(R - \ft12 (\del\phi)^2
   -\ft14 e^{a\phi}\, F_1^2 -\ft14 e^{b\phi}\, F_2^2\Big)\,.
\ee
%%%%%
The theory and its black hole solutions were studied extensively in
$n=d+1$ dimensions in \cite{hong2field}.  It is convenient to 
parameterise the dilaton coupling constants $a$ and $b$ as
%%%%%
\be
a^2= \fft{4}{N_1} - \fft{2(d-2)}{d-1}\,,\qquad
b^2= \fft{4}{N_2} - \fft{2(d-2)}{d-1}\,.
\ee
%%%%%
It was found that while 
black hole solutions with both field strengths carrying charge cannot be
found explicitly for general values of $a$ and $b$, they can be obtained
if 
%%%%%
\be
 a b = -\fft{2(d-2)}{d-1}\,,\label{abrel}
\ee
%%%%%
and we shall assume this from now on.  This condition also implies
%%%%%
\be
a N_1 + b N_2=0\,,\qquad N_1 + N_2=  \fft{2(d-1)}{d-2}\,.
\ee
%%%%%

   The Hamiltonian and Gauss law constraints will be
%%%%%
\bea
   R &=&\ft12 g^{ij}\del_i\phi\, \del_j\phi + \ft12 e^{a\phi}\, g^{ij}\,
         E^1_i E^1_j + \ft12 e^{b\phi}\, g^{ij}\,
         E^2_i E^2_j\,,\label{E2MDHam0}\\
0&=& \nabla_i(e^{a\phi}\, g^{ij}\, E^1_j)\,,\qquad
0=\nabla_i(e^{b\phi}\, g^{ij}\, E^2_j)\label{E2MDGauss}\,.
\eea
%%%%%
We shall again consider a flat Euclidean background metric, with
$g_{ij}= \Phi^{4/(d-2)}\, \delta_{ij}$.
Upon using (\ref{lichR}) the Hamiltonian constraint (\ref{E2MDHam0}) becomes
%%%%%
\be
\fft{4(d-1)}{(d-2)}\, \del_i\del_i \Phi +
  \ft12 \Phi\, \Big[ \del_i\phi\, \del_i\phi + e^{a\phi}\, E^1_i E^1_i
       + e^{b\phi}\, E^2_i E^2_i \Big]=0\,.\label{E2MDHam}
\ee
%%%%%
It is now a straightforward exercise to make an appropriate ansatz for 
solving the Hamiltonian and Gauss law constraints in terms of harmonic
functions, and then to solve for the various exponents in the ansatz in
order to satisfy the constrain equations.  Motivated by the form of the ansatz
that was employed in four dimensions in \cite{cvegibpop}, we have
made an ansatz here involving four harmonic functions, $C_1$, $D_1$, $C_2$ 
and $D_2$, and we find we can solve the Gauss law constraints 
(\ref{E2MDGauss}) and the Hamiltonian constraint (\ref{E2MDHam}) by
writing
%%%%%
\bea
\Phi &=& (C_1\, D_1)^{\ft{(d-2)\,N_1}{4(d-1)}}\, 
         (C_2\, D_2)^{\ft{(d-2)\,N_2}{4(d-1)}}\,,\nn\\
e^\phi&=& (C_1\, D_1)^{\ft12 a N_1}\, (C_2\, D_2)^{\ft12 b N_2}=
   \Big(\fft{C_1\, D_1}{C_2\, D_2}\Big)^{\ft12 a N_1}\,,\nn\\
E^1_i &=& \sqrt{N_1}\, \Big(\fft{C_1\, D_1}{C_2\, D_2}\Big)^{
     -\ft{(d-2)\, N_2}{2(d-1)}}\, \del_i\log\fft{C_1}{D_1}\,,\nn\\
E^2_i &=& \sqrt{N_2}\, \Big(\fft{C_2\, D_2}{C_1\, D_1}\Big)^{
     -\ft{(d-2)\, N_1}{2(d-1)}}\, \del_i\log\fft{C_2}{D_2}\,.\label{datasol}
\eea
%%%%%

\subsection{Static non-extremal black hole}

  The spherically-symmetric non-extremal black hole solutions in the
E2MD theory, when $a$ and $b$ obey the relation (\ref{abrel}), can be
found in \cite{hong2field}.  They are given by
%%%%%
\begin{eqnarray}
ds^2 &=& -(H_1^{N_1} H_2^{N_2})^{-\fft{(D-3)}{D-2}} f dt^2 + 
(H_1^{N_1} H_2^{N_2})^{\fft{1}{D-2}} (f^{-1} dr^2 + r^2 d\Omega_{D-2}^2)\,,\cr
A_1&=& \fft{\sqrt{N_1}\,c_1}{s_1}\, H_1^{-1} dt\,,\qquad
A_2 = \fft{\sqrt{N_2}\,c_2}{s_2}\, H_2^{-1} dt\,,\cr
\phi &=& \ft12 N_1\, a_1 \log H_1 + \ft12 N_2 \,a_2 \log  H_2\,,\qquad
f=1 - \fft{\mu}{r^{D-3}}\,,\cr
H_1&=&1 + \fft{\mu \, s_1^2}{r^{D-3}}\,,\qquad
H_2 = 1 + \fft{\mu \, s_2^2}{r^{D-3}}\,,\label{solution1}
\end{eqnarray}
%%%%%
where we are using a standard notation where $s_i=\sinh\delta_i$ and
$c_i=\cosh\delta_i$.  If we make the coordinate transformation
%%%%%
\be
r^{d-2}= \rho^{d-2}\, \Big(1+ \fft{\mu}{4\rho^{d-2}}\Big)^2\,,
\ee
%%%%%
the metric can be cast into the standard static form (\ref{staticmet}),
with $\Phi$ and $\phi$ 
given as in (\ref{datasol}), where the harmonic functions
take the specific forms
%%%%%
\be
C_1= 1+\fft{\mu\, e^{2\delta_1}}{4\rho^{d-2}}\,,\quad
D_1= 1+\fft{\mu\, e^{-2\delta_1}}{4\rho^{d-2}}\,,\quad
C_2= 1+\fft{\mu\, e^{2\delta_2}}{4\rho^{d-2}}\,,\quad
D_2= 1+\fft{\mu\, e^{-2\delta_2}}{4\rho^{d-2}}\,. \label{specialCD}
\ee
%%%%%
The metric function $g_{00}=-N^2$ is given by
%%%%%
\be
N^2 = \Big(1-\fft{\mu^2}{16\rho^{2(d-2)}}\Big)^2\,
   \Big[(C_1\, D_1)^{N_1}\, (C_2\, D_2)^{N_2}\Big]^{-\ft{d-2}{d-2}}\,.
\ee
%%%%%

   After some calculation, one can verify that the field strengths
in this non-extremal solution indeed imply that $E_i^1$ and $E_i^2$ 
in the initial-value data are consistent with the expressions we found
in (\ref{datasol}) for the general ansatz with four independent
harmonic functions.  Thus we conclude that in the special case
where the four harmonic functions take the particular form given in
(\ref{specialCD}), they give rise to the initial data for the
non-extremal black hole (\ref{solution1}).

\section{Gravity with $p$ Dilatons and $q$ Maxwell Fields}\label{multiemdsec}

\subsection{The theories}

   A general class of theories that encompasses the relevant bosonic
sectors of various supergravities is provided by considering the Lagrangian
%%%%%
\be
{\cal L}= \sqrt{-g}\, \Big( E - \ft12\sum_{\alpha=1}^p (\del\phi_\alpha)^2
  - \ft14 \sum_{I=1}^q X_I^{-2}\,
   F_{(I)}^2\Big)\,, \qquad X_I= e^{-\ft12 \vec a_I\cdot\vec \phi}
\label{genpq}
\ee
%%%%%
in $n=d+1$ spacetime dimensions, 
where $\vec\phi=(\phi_1,\phi_2,\cdots, \phi_p)$ is the $p$-vector of
dilaton fields, and $\vec a_I$ is a set of $q$ constant dilaton $p$-vectors 
that characterise the couplings of the dilatons to the $q$ Maxwell
fields $F_{(I)}$.  We can obtain multi-centre BPS black hole solutions,
and spherically-symmetric static non-extremal black hole solutions, whenever
the dilaton vectors obey the relations \cite{lupomultiscalar}
%%%%%
\be
\vec a_I\cdot\vec a_J = 4 \delta_{IJ} - \fft{2(d-2)}{d-1}\,.\label{adotrel}
\ee
%%%%%
We shall assume the dilaton vectors obey this relation from now on.

   For a given dimension and a given number $p$ of dilaton fields, the most
general theory of the form (\ref{genpq}) will correspond to the case where
$q$ is chosen to be as large as possible, subject to the set of
 dilaton vectors $\vec a_i$ obeying (\ref{adotrel}). Obviously, 
one can always find $p$ such $p$-vectors.  To see this, let $\vec e_I$
be an orthonormal basis in $\R^p$, where the vector $\vec e_i$ has an
entry 1 at the $I$th position, with all other components zero.  Thus
$\vec e_I\cdot\vec e_J=\delta_{IJ}$.  If we define 
$\vec e \equiv \sum_I \vec e_I$ then clearly the vectors
%%%%%
\be
\vec a_I = \alpha \, \vec e_I + \beta\, \vec e
\ee
%%%%%
will obey the relations
%%%%%
\be
\vec a_I\cdot\vec a_J= \alpha^2\, \delta_{IJ} + 2\alpha\beta + p \beta^2\,.
\ee
%%%%%
Solving for $\alpha$ and $\beta$ such that this reproduces (\ref{adotrel}),
we find
%%%%%
\be
   \alpha=2\,,\qquad \beta = -\fft{2}{p} \pm \fft{2}{p}\,
   \sqrt{1-\fft{(d-2)p}{2(d-1)}}\,.
\ee
%%%%%

The only further question
is whether one can find a set of more than these $p$ such vectors,
that all obey (\ref{adotrel}).  Any additional dilaton vector
or vectors, over and above the $p$ already constructed above, 
would necessarily have to be a
linear combination of the first $p$ dilaton vectors.  Let us suppose
that such an additional dilaton vector $\vec u$ existed, over and above
the $p$ dilaton vectors $\vec a_I$ with $1\le I\le p$.  Thus we must
have
%%%%%
\be
\vec u = \sum_{I=1}^p c_I\, \vec a_I\,,
\ee
%%%%%
where $c_I$ are some constants.  The $\vec a_I$ satisfy (\ref{adotrel}), 
and we must also require
%%%%%
\be
\vec u\cdot\vec u= 4-\fft{2(d-2)}{d-1}\,,\qquad
 \vec u\cdot \vec a_I = -\fft{2(d-2)}{d-1}\,.
\ee
%%%%%
Using (\ref{adotrel}) we easily see that these conditions imply
%%%%%
\be
c_I = -1\,,\qquad p=\fft{d}{d-2}\,,
\ee
%%%%%
and so we only have solutions with integer $p$ if 
$(d,p)=(3,3)$ or $(d,p)=(4,2)$.
Thus in four spacetime dimensions we can have a theory of the type
(\ref{genpq}), with three dilatons and four Maxwell fields.  This corresponds 
to the bosonic subsector of four-dimensional STU supergravity in which
the three additional axionic scalars are set to zero.  In five spacetime 
dimensions we can have a theory of the type (\ref{genpq}) with two dilatons
and three Maxwell fields.  This corresponds to a bosonic subsector of
five-dimensional STU supergravity.  In all other cases, the 
requirement that the dilaton vectors in the Lagrangian (\ref{genpq}) obey
the relations (\ref{adotrel}) restricts us to having only $p$ Maxwell fields
when there are $p$ dilatonic scalar fields.

\subsection{Ansatz for initial-value constraints}

   The time-symmetric initial-value constraints for the theory (\ref{genpq}) 
are 
%%%%%
\bea
R &=& \ft12 g^{ij}\, \del_i\vec\phi\cdot \del_i\vec\phi + 
       \ft12 g^{ij}\, \sum_{I=1}^q X_I^{-2}\,E^I_i\, E^I_j\,,\label{Hamilgen}
\\
0 &=& \nabla_i( g^{ij}\, X_I^{-2}\, E^I_j)\,.\label{Gaussgen}
\eea
%%%%% 
We shall, as usual, use a flat background $d$-metric, and so we write
%%%%%
\be 
g_{ij}= \Phi^{\ft{4}{d-2}}\, \delta_{ij}\,.
\ee
%%%%%
The ansatz for the initial-value data for four-dimensional STU supergravity was
discussed in \cite{cvegibpop}; it involved a total of eight arbitrary 
harmonic functions (two for each of the four Maxwell fields).  Multi-centre
BPS black holes were constructed in arbitrary dimensions 
for the theories described by (\ref{genpq}),
with dilaton vectors obeying (\ref{adotrel}), in \cite{lupotrxu}, and these
provide a useful guide for writing an ansatz for initial-value data
in the general case.   One can also construct spherically-symmetric
non-extremal black hole solutions in all dimensions.  These
solutions provide further guidance for writing an ansatz for
 initial-value data in general dimensions.  In particular, we find
that in order to encompass an initial-value formulation for these
non-extremal solutions, we must go beyond the natural-looking generalisation
of the four-dimensional STU supergravity example that would involve $2q$
harmonic functions for the $q$ Maxwell fields.  Namely, we must introduce
one further arbitrary harmonic function, which we shall call $W$.  

After some experimentation, we are led to
consider the following ansatz for the initial data:
%%%%%
\bea
\Phi &=& \Pi^{\ft{d-2}{4(d-1)}}\, W^\gamma\,, 
  \quad \vec\phi = \ft12\sum_{I=1}^q
  \vec a_I\, \log (C_I\, D_I)-\vec a\, \log W\,,\quad
E^I_i = \fft{\Phi^2}{C_I\, D_I}\, \del_i
    \log\fft{C_I}{D_I}\,,\label{IVgen}\nn\\
\Pi&\equiv & \prod_{I=1}^q C_I \, D_I\,,\qquad 
   \gamma\equiv 1-\fft{q(d-2)}{2(d-1)}\,,\qquad
  \vec a \equiv \sum_{I=1}^q\, \vec a_I\,.
\eea
%%%%%
(Note that for four-dimensional STU supergravity we have $d=3$ and
$q=4$, implying $\gamma=0$, and $\vec a=0$, so the additional harmonic function $W$ is absent
in this special case.)

   One can easily verify from the definition of $X_I$ in (\ref{genpq}),
and using the relations (\ref{adotrel}), that the ansatz for $\vec\phi$ 
in (\ref{IVgen}) implies
%%%%%
\be
X_I^{-2}= \Phi^{-4}\, (C_I\, D_I)^2\,,\label{XIsol}
\ee
%%%%%
and then it is easy to see that the Gauss law constraints (\ref{Gaussgen})
are satisfied if the functions $C_I$ and $D_I$ are harmonic in the flat 
background metric,
%%%%%
\be
\del_i\del_i\, C_I=0\,,\qquad \del_i\del_i\, D_I=0\,.\label{CIDIharm}
\ee
%%%%%
Using (\ref{lichR}), we find, upon substituting the ans\"atze (\ref{IVgen})
into the Hamiltonian constraint (\ref{Hamilgen}) that it gives
%%%%%
\be
-\Pi^{-1}\, \del_i\del_i \, \Pi + (\del_i\log\Pi)^2 -
  \fft{4\gamma\, (d-1)}{d-2}\, W^{-1}\, \del_i\del_i\, W =
  \ft12\sum_{I=1}^q \Big\{ [\del_i\log(C_I\, D_I)]^2 +
                         [\del_i\log\fft{C_I}{D_I}]^2\Big\}\,.
\ee
%%%%%
After some algebra, we find that this is indeed satisfied if the
functions $W$, $C_I$ and $D_I$ are harmonic, obeying $\del_i\del_i\, W=0$
and (\ref{CIDIharm}).  Thus
we have established that (\ref{IVgen}) indeed gives a solution of the
initial-value constraints (\ref{Hamilgen}) and (\ref{Gaussgen}), where
$W$, $C_I$ and $D_I$ are arbitrary harmonic functions in the Euclidean 
background metric.

   As mentioned earlier, special cases of the theories we are considering
in this section include the gravity, dilaton and Maxwell-field sectors of 
four-dimensional STU supergravity (with $(d,p,q)=(3,3,4)$) and five-dimensional
STU supergravity (with $(d,p,q)=(4,2,3)$). In both of these cases the constant
$\gamma$ in (\ref{IVgen}) is zero, and so the harmonic function $W$ does
not arise in the initial-data ansatz.  The four-dimensional STU supergravity
case was discussed in \cite{cvegibpop}.  Some other special cases also 
correspond to the gravity, dilaton and Maxwell-field sectors of supergravities.
These include a six-dimensional case with $(d,p,q)=(5,2,2)$ and 
a seven-dimensional case with $(d,p,q)=(6,2,2)$.

\subsection{Extremal multi-centre black holes}

   As was discussed in general in \cite{lupotrxu}, these solutions are
given by 
%%%%%
\bea
ds^2 &=& -H^{-\ft{d-2}{d-1}}\, dt^2 + H^{\ft1{d-1}}\, dx^i\, dx^i\,,\nn\\
\vec\phi &=& \ft12 \sum_{I=1}^q\, \vec a_I\, \log H_I \,,\qquad
 A^I = -H_I^{-1}\, dt\,,
\eea
%%%%%
where the $H_I$ are arbitrary harmonic functions in the Euclidean metric
$dx^i\, dx^i$. Clearly this solution matches with the initial data in
(\ref{IVgen}), in the special case with
%%%%%
\be
C_I=H_I\,,\qquad D_I=1\,,\qquad W =1\,.
\ee
%%%%%

\subsection{Non-extremal spherically-symmetric black holes}

   Non-extremal spherically-symmetric black holes solutions can easily
be found in the 
theory defined by (\ref{genpq}) and (\ref{adotrel}), and they are 
given by
%%%%%
\bea
ds^2 &=& -H^{-\ft{d-2}{d-1}}\, f\, dt^2 + 
    H^{\ft1{d-1}}\, \Big( \fft{dr^2}{f} + r^2 \, d\Omega_{d-1}^2\Big)\,,
\label{nonextpq}\\
\vec\phi &=& \ft12 \sum_{I=1}^q \vec a_I\, \log H_I\,,\qquad
A^I= \big(1- H_I^{-1} \big)\, \coth\delta_I\, dt\,,\qquad
  H_I= 1 + \fft{2m \sinh^2\delta_I}{r^{d-2}}\,.\nn
\eea
%%%%%
Introducing a new radial variable $\rho$ by 
%%%%%
\be
r^{d-2}= \rho^{d-2}\, \Big( 1 + \fft{m}{2\rho^{d-2}}\Big)^2\,,
\ee
%%%%%
we find that the metric $ds^2$ in (\ref{nonextpq}) becomes
%%%%%
\be
ds^2 = -N^2\, dt^2 + \Phi^{\ft{4}{d-2}}\, (d\rho^2 + \rho^2\, d\Omega_{d-1}^2)
\,,
\ee
%%%%%
where $\Phi$ is given in (\ref{IVgen}) with the harmonic functions $C_I$,
$D_I$ and $W$ being given by
%%%%%
\be
C_I = 1+\fft{m e^{2\delta_I}}{2\rho^{d-2}}\,,\qquad
D_I = 1+\fft{m e^{-2\delta_I}}{2\rho^{d-2}}\,,\qquad
 W= 1+\fft{m}{2\rho^{d-2}}\,,\label{nonextharm}
\ee
%%%%%
and 
%%%%%
\be
N^2= \Phi^{-4}\, W^2\, \Big(1-\fft{m}{2\rho^{d-2}}\Big)^2\,.
\ee
%%%%%
The functions $H_I$ in (\ref{nonextpq}) are given by $H_I=W^{-2}\, C_I\, D_I$,
and hence the potentials $A^I$ in the non-extremal solution are simply 
given by
%%%%%
\be
  A^I= \Big(-\fft{1}{C_I} + \fft1{D_I}\Big)\, dt\,,
\ee
%%%%%
and the dilatonic scalars are given by the expression in (\ref{IVgen}).
Thus we see that the non-extremal spherically-symmetric black hole solutions
do indeed have initial data given by (\ref{IVgen}), with the harmonic
functions $C_I$, $D_I$ and $W$ taking the special spherically-symmetric
form (\ref{nonextharm}).

\section{Mapping Einstein-Maxwell-Dilaton to Einstein-Scalar}

  It was observed in \cite{ortin}, and developed further in \cite{cvegibpop},
that the time-symmetric initial data for a system of gravity coupled to 
Maxwell fields and dilatonic scalars can straightforwardly mapped into
the time-symmetric initial data for an extended system of scalar
fields coupled to gravity.  Although \cite{ortin,cvegibpop} discussed this
specifically for four-dimensional spacetimes, the extension to arbitrary
dimensions is immediate.  

\subsection{Mapping of Einstein-Maxwell-Dilaton data}\label{EMDmapsec}

   The mapping can be illustrated by considering the EMD theories
with multiple dilatons and Maxwell fields that we discussed in section
\ref{multiemdsec}.  Making the replacement
%%%%%
\be
E^I_i \longrightarrow X_I^{-1}\, \del_i \psi_I\,,\label{Etopsi}
\ee
%%%%%
the Hamiltonian constraint (\ref{Hamilgen}) becomes
%%%%%
\be
R= \ft12 g^{ij}\, (\del_i\vec\phi\cdot\del_j\vec\phi + 
                   \del_i\psi_I\, \del_j\psi_I)\,,
\ee
%%%%%
which is the same as the Hamiltonian constraint for a system of
free scalar fields $\vec\phi,\psi_I)$ coupled to gravity.  In view of
(\ref{XIsol}), the ansatz for $E^I_i$ in (\ref{IVgen}) becomes
simply
%%%%%
\be
\psi_I= \log\fft{C_I}{D_I}\label{psiCD}
\ee
%%%%%
for the scalar fields $\psi_I$.  Furthermore, under (\ref{Etopsi}) 
the Gauss law constraints (\ref{Gaussgen}) give simply
%%%%%
\be
\del_i\Big( C_I\, D_I\, \del_i\psi_I\Big)=0\,,
\ee
%%%%
and so these are indeed satisfied when $\psi_\AA$ is given by (\ref{psiCD}),
since $C_I$ and $D_I$ are harmonic.

\subsection{General $N$-scalar system coupled to gravity}\label{ESsec}

   If we consider a general system of $N$ scalar fields $\sigma_\AA$, 
$1\le \AA\le N$, coupled to gravity and described by the Lagrangian
%%%%%
\be
{\cal L}= \sqrt{-\hat g} (\hat R - \ft12 \hat g^{\mu\nu}\,\sum_{\AA=1}^N
         \del_\mu\sigma_\AA\, \del_\nu\sigma_\AA)\,,
\ee
%%%%%
then the Hamiltonian constraint for time-symmetric initial data is
$R=\ft12 g^{ij}\, \sum_\AA \del_i\sigma_\AA\, \del_j\sigma_\AA$, and so 
writing $g_{ij}=\Phi^{4/(d-2)}$ is usual the constraint becomes
%%%%%
\be
-\fft{4(d-1)}{d-2}\, \del_i\del_i\Phi = \ft12 \Phi\, 
 \sum_{\AA=1}^N \del_i\sigma_\AA\,\del_i\sigma_\AA\,.\label{Hscalar}
\ee
%%%%%
Following the strategy used in four dimensions in \cite{cvegibpop} we make 
the ansatz
%%%%%
\be
\Phi= \prod_{a=1}^M\, X_a^{n_a} \,,\qquad 
  \sigma_\AA= \sqrt{\fft{8(d-1)}{d-2}}\,
  \sum_{a=1}^M m_a^\AA\, \log X_a\,,\label{Phisig}
\ee
%%%%%
where $X_a$ are a set of $M$ harmonic functions, $\del_i\del_i X_a=0$,
and the constants $n_a$ and $m_a^\AA$ are determined by requiring that
the Hamiltonian constraint (\ref{Hscalar}) be satisfied.  One finds
that this holds if 
%%%%%
\be
  n_a\, n_b + \sum_{\AA=1}^N m_a^\AA\, m_b^\AA = n_a\, \delta_{ab}\,.
\label{mndot0}
\ee
%%%%%
As in \cite{cvegibpop}, by defining the $M$ $(N+1)$-component vectors
%%%%%
\be
{\bf m}_a =(m_a^\AA,n_a)\equiv (m_a^1\,m_a^2,\cdots,m_a^N,n_a)\label{bfm}
\ee
%%%%%
in $\R^{N+1}$, (\ref{mndot0}) becomes
%%%%%
\be
{\bf m}_a\cdot {\bf m_b}= n_a\, \delta_{ab}\,.\label{bfmdot}
\ee
%%%%%
Defining ${\bf Q}_a= {\bf m}_a/\sqrt{n_a}$ one has 
%%%%%
\be
{\bf Q}_a\cdot {\bf Q_b}=\delta_{ab}\,,
\ee
%%%%%
show that there is a one-one mapping between orthonormal $M$-frames in
$\R^{N+1}$ and solutions of the Hamiltonian constraint conditions
(\ref{mndot0}).  This means that we must have $M\le N+1$.

\subsection{Specialisation to the scalar theories from EMD}

   In the mapping from Einstein-Maxwell-Dilaton theories with $p$
dilatons and $q$ Maxwell fields to a purely Einstein-Scalar system with
$(p+q)$ scalar fields, which we described in section (\ref{EMDmapsec}),
the initial-value constraints were solved in terms of the harmonic
functions $C_I$, $D_I$ and $W$.  We may relate this to the general
scalar discussion in section (\ref{ESsec}) by noting that, in an obvious
notation, the harmonic functions $X_a$ and the scalar fields $\sigma_\AA$
are now split as
%%%%%
\be
X_a=\{C_I, D_I, W\}\,,\qquad \sigma_\AA 
=\{\psi_I,\, \vec\phi\, \}\,.\label{XCDW}
\ee
%%%%%
Comparing the expressions for $\Phi$, in eqn (\ref{IVgen}), and 
$\psi_I$, in eqn (\ref{psiCD}), with eqn (\ref{Phisig}) we see that 
for these cases the vectors ${\bf m}_a$ defined in (\ref{bfm}) are given,
following the same notation as for $X_a$ in (\ref{XCDW}), by
%%%%%
\bea
{\bf m}_{{\sst C}_I} &=&\beta\, (\vec e_I,\, \ft12\vec a_I,\, 2\beta)\,,\nn\\
{\bf m}_{{\sst D}_I} &=&\beta\, (-\vec e_I,\, \ft12\vec a_I,\, 2\beta)\,,\nn\\
{\bf m}_w &=& (\vec 0,\, -\beta\, \vec a,\, \gamma)\,,\label{mforemd}
\eea
%%%%%
where $\vec e_I$ is an orthonormal basis for $\R^q$, 
$\beta =\sqrt{\ft{d-2}{8(d-1)}}$, and $\gamma$ and $\vec a$ are 
defined in (\ref{IVgen}). One can easily verify that the vectors
defined in (\ref{mforemd}) indeed satisfy (\ref{bfmdot}).

\section{Wormholes}

\subsection{Wormhole initial data for vacuum Einstein equations}

   It was observed by Misner in the case of four spacetime 
dimensions that if one takes the spatial background metric to be $S^1\times
S^2$, then this can give rise to initial data that generates 
wormhole spacetimes.  One can analogously consider $S^1\times S^{d-1}$
spatial backgrounds for $(d+1)$-dimensional spacetimes.  Taking the
background metric to be 
%%%%%
\be
d\bar s^2 = d\mu^2 + d\sigma^2 + \sin^2\sigma\, d\Omega_{d-2}^2\,,
\label{S1Sd-1}
\ee
%%%%%
then from (\ref{lichR}), the condition for the vanishing of the Ricci 
scalar 
for the spatial metric $ds^2 = \Phi^{4/(d-2)}\, d\bar s^2$ is that $\Phi$
should satisfy
%%%%%
\be
-\bar\square \Phi + \fft{(d-2)^2}{4}\, \Phi=0\,,\label{PhieqnS1Sd-1}
\ee
%%%%%
since the $S^1\times S^{d-1}$ metric (\ref{S1Sd-1}) has Ricci scalar $\bar R=
(d-1)(d-2)$.  One can easily see that a solution for $\Phi$ is given by
%%%%%
\be
\Phi= c^{\ft{d-2}{2}}\, 
\Big(\cosh\mu -\cos\sigma\Big)^{-\fft{d-2}{2}}\,,\label{Phising}
\ee
%%%%%
where $c$ is any constant.
 
 If we take the solution (\ref{Phising}) itself, the metric
$ds^2=\Phi^{4/(d-2)}\, d\bar s^2$ is nothing but the flat Euclidean metric
$ds^2=dx^a dx^a + dz^2$ written in bi-hyperspherical coordinates, with the 
Euclidean coordinates $x^i=(x^a,z)$ given by
%%%%%
\be
x^a= \fft{c\, u^a\, \sin\sigma}{\cosh\mu-\cos\sigma}\,,\qquad
z= \fft{c\, \sinh\mu}{\cosh\mu - \cos\sigma}\,,
\ee
%%%%%
where the $u^a$, constrained by $u^a\, u^a=1$, parameterise points on the
unit $(d-2)$-sphere whose metric is $d\Omega_{d-2}^2=du^a\, du^a$.

    Of course since the metric (\ref{S1Sd-1}) is invariant under 
translations of
the $\mu$ coordinate, and
(\ref{PhieqnS1Sd-1}) is a linear equation, one can
form superpositions to obtain the more general solutions
%%%%%
\be
\Phi=\sum_n\, A_n\, \Big(\cosh(\mu-\mu_n) -\cos\sigma\Big)^{-\fft{d-2}{2}}
\,,\label{PhieqnS1Sd-1gen}
\ee
%%%%%
The the $A_n$ and $\mu_n$ are arbitrary constants.  Since one would like
the wormhole metric to be single-valued and hence periodic in the $\mu$
coordinate on $S^1$, it is appropriate to take $\mu_n= -2 n\, \mu_0$ and
$A_n=a^{\ft{d-2}{2}}$, with the summation in (\ref{PhieqnS1Sd-1gen}) 
being taken over
all the integers, and with $2\mu_0$ being the period of $\mu$:
%%%%%
\be
\Phi(\mu,\sigma)= c^{\ft{d-2}{2}}\, \sum_{n\in Z} 
\Big(\cosh(\mu+2n\, \mu_0) -\cos\sigma\Big)^{-\fft{d-2}{2}}
\,.\label{Phiperiodic}
\ee
%%%%%

   In a natural generalisation of the case of four spacetime dimensions
that was discussed in \cite{Lindquist} and elsewhere, one can easily 
see that if we consider the elementary harmonic function 
%%%%%
\be
\fft1{| \bx + \bd_n|^{d-2}}
\ee
%%%%%%
in the Euclidean space with coordinates $\bx=\{x^1,\ldots x^{d-1}, z\}$, where
$-\bd_n$ is the location of the singularity, with 
%%%%%
\be
\bd_n=\{0,\ldots,0,c\, \coth n\mu_0\}\,,\label{dnlocs}
\ee
%%%%%%
then
%%%%%
\be
\fft1{| \bx + \bd_n|^{d-2}} =
   \fft{(\sinh n \mu_0)^{d-2}}{c^{d-2}}\, 
       (\cosh\mu -\cos\sigma)^{\ft{d-2}{2}}\, \Big(\cosh (\mu+2n \mu_0)
          -\cos\sigma\Big)^{-\ft{d-2}{2}}\,,\label{elterm}
\ee
%%%%%%
and so the periodic conformal function $\Phi$ constructed in (\ref{Phiperiodic})
can be expressed as 
%%%%%
\be
\Phi = c^{\ft{d-2}{2}}\, \Big(\cosh\mu - \cos\sigma\Big)^{-\ft{d-2}{2}}\,
   \hat\Phi\,,\label{PhihatPhi}
\ee
%%%%%
where
%%%%%
\be
\hat\Phi = 1 + \sum_{n\ge 1}\fft{c^{d-2}}{(\sinh n \mu_0)^{d-2}}\, 
\Big[\, \fft1{|\bx + \bd_n|^{d-2}} + 
   \fft1{|\bx - \bd_n|^{d-2}} \,\Big]\,.
\ee
%%%%%
Thus the initial-time spatial $d$-metric  
metric $ds^2=\Phi^{\ft4{d-2}}\, d\bar s^2$, where $d\bar s^2$ is
the $S^1\times S^{d-1}$ metric (\ref{S1Sd-1}), can be written as
%%%%%%
\be
ds^2 = \hat\Phi^{\ft{4}{d-2}}\, dx^i dx^i\,,
\ee
%%%%%%
which is the metric for a sum over infinitely-many mass points at the 
locations $-\bd_n$ and $+\bd_n$, with strengths $c^{d-2}\, 
(\sinh n\mu_0)^{-(d-2)}$, giving a description of a two-centre 
wormhole in $(d+1)$ spacetime dimensions.  Comparing with the form of
the multi-black hole initial data discussed in section \ref{vacuumflatsec},
we see that the total mass of the wormhole is given by
%%%%%
\be
M= 4 c^{d-2}\, \sum_{p\ge 1} \fft1{(\sinh n\mu_0)^{d-2}}\,.\label{whmass}
\ee
%%%%%
This generalises the result for the four-dimensional spacetime wormhole
considered by Misner in \cite{Misner:1960zz}, which corresponded to the
case $d=3$.

   The infinite sums in the expression (\ref{whmass}) can in fact be evaluated
explicitly, in terms of the $q$-polygamma function 
%%%%%
\be
\psi^{(r)}_q(z)= \fft{\del^r\, \psi_q(z)}{\del z^r}\,,\qquad
\psi_q(z)=\fft1{\Gamma_q(z)}\, \fft{\del \Gamma_q(z)}{\del z}=
-\log(1-q) +\log q\, \sum_{k\ge 0}\fft{q^{k+z}}{1-q^{k+z}}\,,
\ee
%%%%%
where $\Gamma_q(z)$ is the $q$ generalisation of the usual gamma function
$\Gamma(z)$.  For example, one finds that
%%%%%
\bea
\sum_{n\ge 1}\fft1{\sinh n \mu_0} &=& \fft{\im \pi - \psi_q(1) 
           + \psi_q(1-\ft{\im \pi}{\mu_0})}{\mu_0}\,,\nn\\
\sum_{n\ge 1}\fft1{(\sinh n \mu_0)^2} &=&
  \fft{-2\mu_0 + \psi^{(1)}_q(1) + \psi^{(1)}_q(1-\ft{\im\pi}{\mu_0})}{
            \mu_0^2}\,,
\eea
%%%%%
where $q\equiv e^{\mu_0}$.  Interestingly, in the special case $\mu_0=\pi$,
the sum for $d=4$ has a simple expression,
%%%%%
\be
\sum_{n\ge 1}\fft1{(\sinh n \pi)^2} = \fft1{6}-\fft1{2\pi}\,.
\ee
%%%%%
Another example with a simple expression is when $\mu_0=\pi$ in $d=6$, for
which one has
%%%%%
\be
\sum_{n\ge 1}\fft1{(\sinh n \pi)^4} = \fft1{3\pi} -\fft{11}{90} +
    \fft{[\Gamma(\ft14)]^8}{1920 \pi^6}\,.
\ee
%%%%%

\subsection{Wormhole initial data for Einstein-Maxwell}

   By an elementary extension of the calculation described in section
\ref{EMsec}, one can verify that writing  
%%%%%
\be
ds^2 = \Phi^{\ft4{d-2}}\, d\bar s^2\,,
\ee
%%%%%
where $d\bar s^2$ is the metric (\ref{S1Sd-1}) on $S^1\times S^{d-1}$, the
initial value constraints (\ref{Hamconeinmax}) and $\nabla_i E^i=0$
and for the Einstein-Maxwell system are satisfied by again writing
%%%%%
\be
\Phi=(C D)^{\ft12}\,,\qquad E_i= \sqrt{\fft{d-1}{2(d-2)}}\, \del_i
   \log\fft{C}{D}\,,\label{PhiCD2}
\ee
%%%%%
where now $C$ and $D$ are arbitrary solutions of the Helmholtz equation
%%%%%
\be
-\bar\square C+ \fft{(d-2)^2}{4}\, C=0\,,\qquad 
  -\bar\square D +  \fft{(d-2)^2}{4}\, D=0\,.
\ee
%%%%%
Thus we can solve the constraint equations by taking each of $C$ and $D$ 
to be functions of the general form (\ref{PhieqnS1Sd-1gen}).  Since
we would again like to construct initial data that is periodic in the
circle coordinate $\mu$, it is important that we have $\Phi(\mu,\sigma)=
\Phi(\mu+2\mu_0,\sigma)$, where $2\mu_0$ is the period of $\mu$.  However,
as can be seen from (\ref{PhiCD2}), the functions $C$ and $D$ can be allowed
to have the more general holonomy properties
%%%%%
\be
C(\mu+2\mu_0,\sigma)= e^{-\lambda}\, C(\mu,\sigma)\,,\qquad
D(\mu+2\mu_0,\sigma)= e^{\lambda}\, D(\mu,\sigma)\,,
\ee
%%%%%
where $\lambda$ is a constant.  This is compatible also with the 
single-valuedness of the solution for $E_i$ in (\ref{PhiCD2}).  We
can construct solutions $C$ and $D$ with the required holonomy by 
taking
%%%%%
\bea
C(\mu,\sigma) &=& c^{\ft{d-2}{2}}\, \sum_{n\in Z} e^{n\lambda}\,
   \Big(\cosh(\mu+2n\mu_0) - \cos\sigma\Big)^{-\ft{d-2}{2}}\,,\nn\\
D(\mu,\sigma)&=&c^{\ft{d-2}{2}}\, \sum_{n\in Z} e^{-n\lambda}\,
   \Big(\cosh(\mu+2n\mu_0) - \cos\sigma\Big)^{-\ft{d-2}{2}}\,.\label{CDhol}
\eea
%%%%%
These series are convergent provided that $|\lambda|<(d-2)\, |\mu_0|$.

  It is again useful to re-express $C$ and $D$ in terms of 
harmonic functions $\hat C$ and $\hat D$ in the conformally-related 
Euclidean space.  Thus from (\ref{elterm}) we see that $\hat C$ and $\hat D$
defined by
%%%%%
\be
C= \fft{c^{\ft{d-2}{2}}}{[\cosh\mu -\cos\sigma]^{\ft{d-2}{2}}}\,\hat C\,,\qquad
D= \fft{c^{\ft{d-2}{2}}}{[\cosh\mu -\cos\sigma]^{\ft{d-2}{2}}}\,\hat D
\label{CDhChD}
\ee
%%%%%
are given by
%%%%%
\bea
\hat C &=& 1 + \sum_{n\ge 1}\, \fft{c^{d-2}}{(\sinh n\mu_0)^{d-2}}\,
 \Big[ \fft{e^{n\lambda}}{|\bx + \bd_n|^{d-2}} +
   \fft{e^{-n\lambda}}{|\bx - \bd_n|^{d-2}}\Big]\,,\nn\\
\hat D &=& 1 + \sum_{n\ge 1}\, \fft{c^{d-2}}{(\sinh n\mu_0)^{d-2}}\,
 \Big[ \fft{e^{-n\lambda}}{|\bx + \bd_n|^{d-2}} +
   \fft{e^{n\lambda}}{|\bx - \bd_n|^{d-2}}\Big]\,.\label{hChDdef}
\eea
%%%%%
Defining $\hat\Phi=(\hat C \hat D)^{\ft12}$ we therefore have
%%%%%
\be
ds^2 = \Phi^{\ft{4}{d-2}}\, d\bar s^2 = \hat\Phi^{\ft{4}{d-2}}\,
  dx^i dx^i\,,\label{metricgen}
\ee
%%%%%
and so we straightforwardly find that the total mass of the Einstein-Maxwell
wormhole is given by
%%%%%
\be
M= 4 c^{d-2}\, \sum_{n\ge 1} \fft{\cosh n\lambda}{(\sinh n\mu_0)^{d-2}}\,.
\label{MEM}
\ee
%%%%%
This reduces to the result in \cite{Lindquist} when $d=3$, 
corresponding to the case of Einstein-Maxwell wormholes in four
spacetime dimensions.  The mass is finite provided that the condition
 $|\lambda|<  (d-2)\, |\mu_0|$ that we mentioned previously is satisfied.

  The electric charge threading each wormhole throat can be calculated from a
Gaussian integral
%%%%%
\be
Q= \fft1{\omega_{d-1}}\, \int_S E_i\, n^i\, dS\,,
\ee
%%%%%
where $\omega_{d-1}$ is the volume of the unit $(d-1)$ sphere, and $n^i$ is the
unit vector normal to the $(d-1)$-surface with area element $dS$ enclosing
the charged mass points that comprise the wormhole throat under 
consideration.  In our case the mass points at $\bd_n$ for $1\le n\le\infty$ 
are associated with one throat, and the mass points at $-\bd_n$ with the
other.
Since the spatial metric $ds^2$ on the initial surface is equal to
$(\hat C\hat D)^{\ft{2}{d-2}}\, dx^i dx^i$, the area element $dS =
(\hat C\hat D)^{\ft{d-1}{d-2}}\, d\hat S$ and the unit vector 
$n^i = (\hat C\hat D)^{-\ft1{d-2}}\, \hat n^i$, where $d\hat S$ and $\hat n^i$
are the corresponding quantities in the Euclidean metric $dx^i dx^i$.  
Thus, from (\ref{PhiCD2}) and (\ref{CDhChD}) we have
%%%%%
\be
Q= \fft{1}{\omega_{d-1}}\, \sqrt{\fft{d-1}{2(d-2)}}\, 
  \int_{\hat S}\,(\hat D\, \del_i \hat C- \hat C\, \del_i \hat D)\,
   d\hat S^i\,.\label{QCD}
\ee
%%%%%
The corresponding charge for the other throat will be $-Q$.

   One way to evaluate (\ref{QCD}) is to convert it, using the divergence
theorem, into 
\be
Q= \fft{1}{\omega_{d-1}}\, \sqrt{\fft{d-1}{2(d-2)}}\,
  \int_{V}\,(\hat D\, \del_i\del_i \hat C- \hat C\, \del_i \del_i \hat D)\,
   d^dx\,,\label{QCD2}
\ee
%%%%%
and make use of the fact that the harmonic functions $\hat C$ and $\hat D$,
defined in (\ref{hChDdef}), satisfy
%%%%%
\bea
\del_i\del_i \hat C &=& -(d-2) 
\sum_{n\ge 1}\, \fft{c^{d-2}\, \omega_{d-1}}{(\sinh n\mu_0)^{d-2}}\,
 [e^{n\lambda}\, \delta^d(\bx+\bd_n)  +
    e^{-n\lambda}\, \delta^d(\bx-\bd_n)]\,,\nn\\
\del_i\del_i \hat D &=& -(d-2)
\sum_{n\ge 1}\, \fft{c^{d-2}\, \omega_{d-1} }{(\sinh n\mu_0)^{d-2}}\,
 [e^{-n\lambda}\, \delta^d(\bx+\bd_n)  +
    e^{n\lambda}\, \delta^d(\bx-\bd_n)]\,.
\eea
%%%%%

    To calculate the total charge for the wormhole throat corresponding to
the $\bx = \bd_n$ sequence of mass points, we should choose the
volume $V$ in integral (\ref{QCD2}) to enclose all these mass points, but
none of those located at the points $\bx=-\bd_n$.  Thus we find
%%%%%
\be
Q= \sqrt{\fft{(d-1)(d-2)}{2}}\,\Big[ 
\sum_{n\ge 1} \fft{2 c^{d-2}\,\sinh n\lambda}{
   (\sinh(n\mu_0)^{d-2}} + 
\sum_{m\ge1}\sum_{n\ge 1}\, \fft{2 c^{2d-4}\, \sinh(m+n)\lambda}{
|\bd_m+\bd_n|\, (\sinh m \mu_0\, \sinh n\mu_0)^{d-2}}\Big]\,.
\label{Qsum}
\ee
%%%%%%
Note that the terms that arise 
involving $|\bd_m - \bd_n|^{-(d-2)}$ cancel by 
antisymmetry.   The $m=n$ ``self-energy'' terms require some care, since the 
denominators $|\bd_m - \bd_n|^{d-2}$
go to zero.  One way to handle this is to introduce regulators by sending
$\bx\rightarrow \bx +\bm{\epsilon}$ in $\hat C$, and $\bx\rightarrow 
\bx -\bm{\epsilon}$ in $\hat D$.  The terms involving 
$|\bd_m - \bd_n|^{-(d-2)}$ now become 
$|\bd_m - \bd_n+2 \bm{\epsilon}|^{-(d-2)}$ and still cancel by 
antisymmetry, prior to sending the regulator to zero.  An alternative way to 
evaluate the charge is to work directly with the expression (\ref{QCD}) for
$Q$, and evaluate the contribution for each of the included mass points
$\bx=\bd_n$ by integrating over a small $(d-1)$-sphere
surrounding that point.  After summing over the contributions from all
the mass points, one arrives at the same result (\ref{Qsum}) that we obtained 
above.  In this calculation, the analogous regularisation of the 
potentially-divergent ``self-energy'' terms occurs because they cancel
pairwise by antisymmetry before taking the limit in which the radius of the
small spheres surrounding the mass points goes to zero.

In view of the definition (\ref{dnlocs}) for $\bd_n$ we
have $|\bd_m+\bd_n|= c\, (\coth m\mu_0 + \coth n\mu_0)$, and hence
%%%%%
\bea
Q &=& \sqrt{\fft{(d-1)(d-2)}{2}}\, \sum_{n\ge 1} \fft{2 c^{d-2}\,\sinh n\lambda}{
	(\sinh(n\mu_0)^{d-2}} + \sqrt{\fft{(d-1)(d-2)}{2}}\,
\sum_{m\ge1}\sum_{n\ge 1}\, \fft{2 c^{d-2}\, \sinh(m+n)\lambda}{
 (\sinh (m+n) \mu_0 )^{d-2}}\,,\nn\\
 &=& \sqrt{\fft{(d-1)(d-2)}{2}}\,
\sum_{m\ge0}\sum_{n\ge 1}\, \fft{2 c^{d-2}\, \sinh(m+n)\lambda}{
 (\sinh (m+n) \mu_0 )^{d-2}}\,.\label{Qres2}
\eea
%%%%%%
Defining $p=m+n$, the double summation $\sum_{m=0}^\infty \sum_{n=1}^\infty$
can be rewritten as $\sum_{p\ge 1} \sum_{n=1}^p$, and so (\ref{Qres2})
becomes\footnote{This agrees in the special case $d=3$ with what Lindquist 
would have had in his result for four-dimensional spacetime, if he
had not accidentally omitted the factor of $p$ in the numerator of his
expression.}
%%%%%%
\be
Q=2 c^{d-2}\, 
    \sqrt{\fft{(d-1)(d-2)}{2}}\, \sum_{p\ge 1} \fft{p\, \sinh p\lambda}{
	    (\sinh p\mu_0)^{d-2}}\,.\label{QEM}
\ee
%%%%%%

   It is interesting to note from the expressions (\ref{MEM}) for the
total mass $M$ and (\ref{QEM}) for the charge that
%%%%%
\be
Q= \fft12 \sqrt{\fft{(d-1)(d-2)}{2}}\, \fft{\del M}{\del\lambda}\,.
\ee
%%%%%

\subsection{Einstein-Maxwell-Dilaton wormholes}

  The Einstein-Maxwell-Dilaton system discussed in section \ref{EMDsec} allows
one to set up wormhole initial data also.  One can straightforwardly check
that by taking the background metric $d\bar s^2$ to be the $S^1\times S^{d-1}$
metric (\ref{S1Sd-1}), the Hamilton and Gauss-law constraints (\ref{emdcons})
are satisfied if $\Phi$, $\phi$ and $E_i$ are given by (\ref{emddata})
and the functions $C$, $D$ and $W$ now obey the Helmholtz equations
%%%%%
\be
-\bar\square C+ \fft{(d-2)^2}{4}\, C=0\,,\quad 
  -\bar\square D +  \fft{(d-2)^2}{4}\, D=0\,,\quad
  -\bar\square W +  \fft{(d-2)^2}{4}\, W=0\,.\label{CDWeqns}
\ee
%%%%%
As in the previous wormhole examples, we can construct solutions 
with appropriate periodicity or holonomy properties by taking 
suitable linear superpositions of elementary solutions. In this example,
we see that we can ensure the necessary periodicity of the conformal
factor $\Phi$, the dilaton $\phi$ and the electric field $E_i$ by
arranging that the solutions for $C$, $D$ and $W$ obey\footnote{Note
that we do not encounter the problems that were seen in \cite{ortin} for
the Einstein-Maxwell-Dilaton system, where the dilaton had non-trivial
monodromy and was not periodic in the $\mu$ coordinate.  This is related to
the fact that our ansatz involves three harmonic functions, $C$, $D$ and $W$,
thus generalising the 3-function ansatz in \cite{cvegibpop}, whereas the
more restrictive ansatz in \cite{ortin} has only two harmonic functions.}
%%%%%
\be
C(\mu+ 2\mu_0,\sigma)= e^{-\lambda}\, C(\mu,\sigma)\,,\quad
D(\mu+ 2\mu_0,\sigma)= e^{\lambda}\, D(\mu,\sigma)\,,\quad
W(\mu+2\mu_0,\sigma)= W(\mu,\sigma)\,,
\ee
%%%%%
and so we may take
%%%%%
\bea
C(\mu,\sigma) &=& c^{\ft{d-2}{2}}\, \sum_{n\in Z} e^{n\lambda}\,
   \Big(\cosh(\mu+2n\mu_0) - \cos\sigma\Big)^{-\ft{d-2}{2}}\,,\nn\\
D(\mu,\sigma)&=&c^{\ft{d-2}{2}}\, \sum_{n\in Z} e^{-n\lambda}\,
   \Big(\cosh(\mu+2n\mu_0) - \cos\sigma\Big)^{-\ft{d-2}{2}}\,,\nn\\
W(\mu,\sigma)&=&c^{\ft{d-2}{2}}\, \sum_{n\in Z}
   \Big(\cosh(\mu+2n\mu_0) - \cos\sigma\Big)^{-\ft{d-2}{2}}\,.
\label{CDWhol}
\eea
%%%%%
We again have the expressions (\ref{metricgen}) for the metric $ds^2$ in 
terms of the $S^1\times S^{d-1}$ metric $d\bar s^2$ and the Euclidean metric
$dx^i dx^i$, with $\hat\Phi$ related to $\Phi$ as in (\ref{PhihatPhi}) and
now
%%%%%
\be
\hat\Phi = (\hat C\, \hat D)^{\ft{(d-2)}{(d-1)\Delta}}\, 
\hat W^{\ft{a^2}{\Delta}}\,,
\ee
%%%%%
with 
%%%%%
\bea
\hat C &=& 1 + \sum_{n\ge 1}\, \fft{c^{d-2}}{(\sinh n\mu_0)^{d-2}}\,
 \Big[ \fft{e^{n\lambda}}{|\bx + \bd_n|^{d-2}} +
   \fft{e^{-n\lambda}}{|\bx - \bd_n|^{d-2}}\Big]\,,\nn\\
\hat D &=& 1 + \sum_{n\ge 1}\, \fft{c^{d-2}}{(\sinh n\mu_0)^{d-2}}\,
 \Big[ \fft{e^{-n\lambda}}{|\bx + \bd_n|^{d-2}} +
   \fft{e^{n\lambda}}{|\bx - \bd_n|^{d-2}}\Big]\,,\label{hChDhWdef}\\
\hat W &=&1 + \sum_{n\ge 1}\, \fft{c^{d-2}}{(\sinh n\mu_0)^{d-2}}\,
 \Big[ \fft{1}{|\bx + \bd_n|^{d-2}} +
   \fft{1}{|\bx - \bd_n|^{d-2}}\Big]\,.
\eea
%%%%%
Comparing with the form of $\Phi$ in section \ref{vacuumflatsec} we 
straightforwardly find that the total wormhole mass is given by
%%%%%
\be
M= \fft{16(d-2)\, c^{d-2}}{(d-1)\Delta}\, \sum_{n\ge1} 
    \fft{\sinh^2\ft{n\lambda}{2}}{(\sinh(n\mu_0)^{d-2}} +
       4c^{d-2}\, \sum_{n\ge1} \fft1{(\sinh n\mu_0)^{d-2}}\,.\label{MEMD}
\ee
%%%%%%

  The calculation of the electric charge proceeds in a very similar fashion
that for the Einstein-Maxwell case, which we discussed earlier.  Now,
we shall have
%%%%%
\bea
Q &=& \fft1{\omega_{d-1}}\, \int e^{a\phi}\, E_i\, n^i\, dS\,,\nn\\
&=& \fft{2}{\sqrt{\Delta}\, \omega_{d-1}}\, 
   \int(\hat D\, \del_i\hat C - \hat C\, \del_i \hat D)\, d\hat S^i\,,
\eea
%%%%%
and hence by the same steps as for Einstein-Maxwell, we find
%%%%%
\be
Q= \fft{4(d-2) c^{d-2}}{\sqrt{\Delta}}\, 
     \sum_{p\ge 1} \fft{p \sinh p\lambda}{(\sinh p\mu_0)^{d-2}}\,.\label{QEMD}
\ee
%%%%%
We again have a simple relation between the mass $M$, given by (\ref{MEMD}), 
and the charge $Q$ given by (\ref{QEMD}), namely
%%%%%
\be
Q= \fft{(d-1)\, \sqrt{\Delta}}{2}\, \fft{\del M}{\del\lambda}\,.
\ee
%%%%%

\subsection{Multi-Maxwell wormholes}

  For the remaining examples of time-symmetric initial data involving
multiple Maxwell fields, which we discussed in sections \ref{E2MDsec}
and \ref{multiemdsec}, we shall just briefly summarise the results for
wormhole initial data.  

   In the case of two Maxwell fields and a single
dilaton, described in section \ref{E2MDsec}, we find that
each of the pairs of functions $(C_1,D_1)$ and $(C_2,D_2)$ will now take
the form given in (\ref{CDhol}), with independent $\lambda$ 
parameters $\lambda_1$ and $\lambda_2$ allowed for the two pairs, so that
%%%%%
\be
C_I(\mu,\sigma)=e^{-\lambda_I}\, C_I(\mu+2\mu_0,\sigma)\,,\qquad
D_I(\mu,\sigma)=e^{\lambda_I}\, D_I(\mu+2\mu_0,\sigma)\,,\quad I=1,2\,.
\ee
%%%%%
The total wormhole mass is then given by
%%%%%
\be
M= \fft{2(d-2)\, c^{d-2}}{(d-1)}\, \sum_{n\ge 1} 
    \fft{1}{(\sinh n\mu_0)^{d-2}}\,
\Big[ N_1\, \cosh n\lambda_1 + N_2\, \cosh n\lambda_2\Big]\,.
\ee
%%%%%
There are now two charges, one for each Maxwell field, and these are
given by
%%%%%%
\be
Q_I= 2(d-2) c^{d-2}\, \sqrt{N_I}\, \sum_{p\ge 1}
    \fft{p\, \sinh p\lambda_I}{(\sinh p \mu_0)^{d-2}}\,.
\ee
%%%%%
The charges can be expressed in terms of the mass as follows:
%%%%%
\be
Q_I= \fft{(d-1)}{\sqrt{N_I}}\, \fft{\del M}{\del\lambda_I}\,.
\ee
%%%%%

     For the case of $p$ dilaton fields and $q$ Maxwell fields 
discussed in section \ref{multiemdsec}, we find that the initial-value
constraints can be solved, in the $S^1\times S^{d-1}$ background metric, 
by the ansatz (\ref{IVgen}), where now the functions $C_I$, $D_I$ and
$W$ obey the equation (\ref{CDWeqns}).  Single-valuedness of the metric,
dilatons and electric fields requires that we have the holonomy
relations
%%%%%
\be
C_I(\mu+ 2\mu_0,\sigma)= e^{-\lambda_I}\, C_I(\mu,\sigma)\,,\quad
D_I(\mu+ 2\mu_0,\sigma)= e^{\lambda_I}\, D_I(\mu,\sigma)\,,\quad
W(\mu+2\mu_0,\sigma)= W(\mu,\sigma)\,.
\ee
%%%%%
We can take the functions $C_I$, $D_I$ and $W$ to be given as in 
(\ref{CDWhol}), with the different $\lambda_I$ parameters for each
pair $(C_I,D_I)$.   We find the total mass of the wormhole is given by
%%%%%
\be
M= \fft{4(d-2)\, c^{d-2}}{(d-1)}\, \sum_{I=1}^q \sum_{n\ge1}
    \fft{\sinh^2\ft{n\lambda_I}{2}}{(\sinh(n\mu_0)^{d-2}} +
       4c^{d-2}\, \sum_{n\ge1} \fft1{(\sinh n\mu_0)^{d-2}}\,.
\ee
%%%%%%
The total 
charges associated with one of the two wormhole throats are given by
%%%%%
\be
Q_I = 2(d-2) c^{d-2}\, \sum_{p\ge 1} 
   \fft{p\,\sinh p\lambda_I}{(\sinh p \mu_0)^{d-2}}\,,
\ee
%%%%%
with the other carrying charges of equal magnitudes but opposite signs.
The charges and the mass are related by
%%%%%
\be
Q_I= (d-1)\, \fft{\del M}{\del\lambda_I}\,.
\ee
%%%%%

\subsection{Wormhole interaction energy}

A manifold with $N$ Einstein-Rosen bridges in an asymptotically flat spacetime,
with each bridge leading to a different 
asymptotically flat spacetime, has a metric of the form \cite{Brill:1963yv}
%%%%%
\be
ds^2 =  \hat\Phi^{\ft{4}{d-2}}\,dx^i dx^i\,,
\ee
%%%%%
where 
%%%%%
\be
\hat\Phi= 1+ \sum_{i=1}^N \fft{\alpha_n}{r_i^{d-2}}\,,
\ee
%%%%%
and $r_i=|\bx -\bx_i|$, with $\bx_i$ being the location of the $i$th mass point.
The total mass $M$ of this system is given by 
%%%%%
\be
M= 2\sum_{i=1}^N \alpha_i\,.
\ee
%%%%%

In the limit when $\bx$ approaches the $i$th mass point, one has
%%%%%
\be
r_i \rightarrow 0, \qquad r_{j} \rightarrow r_{ij} \; \; (j \neq i)\,,
\ee
%%%%%
and the 
metric takes the form 
%%%%%
\be
ds^2 \rightarrow \left[ \fft{\alpha_i}{r_n^{d-2}} + 
A_n \right]^\ft{4}{d-2} (dr_i^2 + r_i^2 \; d\Omega_{d-1}^2) = 
\left( \fft{\alpha_i}{r_i^{d-2}} \right)^\ft{4}{d-2}  
\left[ 1 + A_i \fft{r_n^{d-2}}{\alpha_n} \right]^\ft{4}{d-2} 
(dr_i^2 + r_i^2 \; d\Omega_{d-1}^2),
\ee
%%%%%
where
%%%%% 
\be
A_i = 1 + \sum_{j \neq i} \fft{\alpha_j}{r_{ij}^{d-2}} \,,
\ee
%%%%%
and we are using $r_i$ as the radial coordinate near $r_i=0$.
Now introducing the new coordinate
%%%%% 
\be
r_i'^{d-2} = \fft{\alpha_i^2}{r_i^{d-2}}\,,
\ee
%%%%%
the line element in the corresponding limit 
$r_i' \rightarrow \infty$ takes the form 
%%%%%
\be
ds^2 \rightarrow \left[ 1 +  \fft{A_i \alpha_i}{r_i'^{d-2}} 
\right]^\ft{4}{d-2} (dr_i'^2 + r_i'^2 \; d\Omega_{d-1}^2).
\ee
%%%%%
This implies that the bare mass of the individual bridge is 
%%%%%
\be
m_i = 2A_i \,\alpha_i = 2\alpha_i + 2\alpha_i \, 
\sum_{j \neq i} \fft{\alpha_j}{r_{ij}^{d-2}}\,,
\ee
%%%%%
and their sum,
%%%%%
\be
\sum_{i=1}^N m_i = 2 \, \sum_{i=1} \alpha_i + 
  2 \sum_{i=1}^N  \sum_{j \neq i} \fft{\alpha_i \, 
\alpha_j}{r_{ij}^{d-2}} \, = \, M + 2 \, \sum_{i=1}^N  \sum_{j \neq i} 
\fft{\alpha_i \, \alpha_j}{r_{ij}^{d-2}}\,,  
\ee
%%%%%
is not equal to the total mass of the system. Hence, the energy of 
gravitational interaction is
%%%%%
\be
M_{int} = M - \sum_{i=1}^N \, m_i = -  2 \, \sum_{i=1}^N  
\sum_{j \neq i} \, \fft{\alpha_i \, \alpha_j}{r_{ij}^{d-2}}. 
\ee
We can now use these results to obtain the 
interaction energy of various wormholes.

\subsubsection{Einstein wormhole}

For the wormhole manifold in pure Einstein gravity we have 
%%%%%
\be
\hat\Phi = 1 + \sum_{n\ge 1}\fft{c^{d-2}}{(\sinh n \mu_0)^{d-2}}\, 
\Big[\, \fft1{|\bx + \bd_n|^{d-2}} + 
\fft1{|\bx - \bd_n|^{d-2}} \,\Big]\,.
\ee
%%%%%
All image points at $\bd_n$ contribute to the mass $m_1$ of one mouth 
of the wormhole, and the rest (at $-\bd_n$) to the mass $m_2$ of the 
other mouth. 
The mass of the $n^{th}$ image point is 
%%%%%
\be
{\mathfrak m}_n = 2\alpha_n + 2\alpha_n \, \sum_{m \neq n} \fft{\alpha_m}{r_{nm}^{d-2}},
\ee 
where
%%%%%
\be
\alpha_n = \fft{c^{d-2}}{(\sinh n \mu_0)^{d-2}}.
\ee
%%%%%
Hence, the masses of the wormhole mouths are
\be
m_2 = m_1 = \sum_{n \ge 1} {\mathfrak m}_n = \fft{M}{2} + 2 \sum_{n \ge 1}  
\sum_{m \neq n} \, \fft{\alpha_n \, \alpha_m}{r_{mn}^{d-2}}\,.\label{withinf}
\ee
%%%%%
The terms where $m$ is negative will have denominators $r_{mn}=|\bd_m -\bd_n|$
of the form $|\bd_p+\bd_n|= c \sinh(p+n) mu_0$, with $n$ and $p=-m$ both 
positive.  The double
sum will converge for such terms.  However, when $m$ is positive the
denominators will be $|\bd_m-\bd_n|= c |\sinh(m-n)\mu_0|$ with $m$ and $n$ 
positive, and even though the terms with $m=n$ are excluded, the
double sum will diverge.  As discussed in detail in \cite{Lindquist}, 
this problem can be resolved by subtracting out the (infinite) interaction
energy between the bare masses which together make up $m_i$.  In other 
words, one makes a ``mass renormalisation'' by adding the infinite
(negative) term
%%%%%
\be
\delta m_1 = 
-2 \sum_{n\ge 1} \sum_{m\ge1,\, m\ne n} \fft{\alpha_n\, \alpha_m}{
r_{mn}^{d-2}} = 
-2 \sum_{n\ge 1} \sum_{\substack{m\ge1\\ m\ne n}} \fft{\alpha_n\, \alpha_m}{
  |\bd_m-\bd_n|^{d-2}} \,.
\ee
%%%%%
This leads to the 
``renormalised'' mass
%%%%%
\be
m_2 = m_1  = \fft{M}{2} + 2 \sum_{n \ge 1}  \sum_{m \leq 1} \, \fft{\alpha_n \, \alpha_m}{r_{mn}^{d-2}} = \fft{M}{2} + 2  \sum_{n \ge 1}  \sum_{m \geq 1} \, \fft{c^{d-2}}{[\sinh(m+n)\mu_0]^{d-2}},
\ee
or, after reorganising the double summation,  
\be
m_2 = m_1 = \fft{M}{2} + 2 \,c^{d-2}\,\sum_{p \ge 2}  
  \fft{p-1}{[\sinh p\mu_0]^{d-2}}.
\ee
Now, the (finite) interaction energy between the two mouths is given by
\be
M_{int} = M - (m_1 + m_2) = 
  - 4 \,c^{d-2}\,\sum_{p \ge 2}  \fft{p-1}{[\sinh p\mu_0]^{d-2}}.
\ee

\subsubsection{Other wormholes}
 
   We may now apply the same procedure to the case of $q$ Maxwell and $p$ 
dilaton fields. For this system we have 
%%%%%
\bea
\hat \Phi = \hat \Pi^{\ft{d-2}{4(d-1)}}\, \hat W^\gamma\,, \qquad
\hat \Pi&\equiv & \prod_{I=1}^q \hat C_I \, \hat D_I\,,\qquad 
\gamma\equiv 1-\fft{q(d-2)}{2(d-1)}\,,
\eea
%%%%%
where 
\bea
\hat C_I &=& 1 + \sum_{n\ge 1}\, \fft{c^{d-2}}{(\sinh n\mu_0)^{d-2}}\,
\Big[ \fft{e^{n\lambda_I}}{|\bx + \bd_n|^{d-2}} +
\fft{e^{-n\lambda_I}}{|\bx - \bd_n|^{d-2}}\Big] = 1+ \sum_{n\neq 0} \fft{c_{In}}{r_n^{d-2}} \, ,\nn\\
\hat D_I &=& 1 + \sum_{n\ge 1}\, \fft{c^{d-2}}{(\sinh n\mu_0)^{d-2}}\,
\Big[ \fft{e^{-n\lambda_I}}{|\bx + \bd_n|^{d-2}} +
\fft{e^{n\lambda_I}}{|\bx - \bd_n|^{d-2}}\Big] = 1+ \sum_{n\neq 0} \fft{d_{In}}{r_n^{d-2}}\,,\nn \\
\hat W &=&1 + \sum_{n\ge 1}\, \fft{c^{d-2}}{(\sinh n\mu_0)^{d-2}}\,
\Big[ \fft{1}{|\bx + \bd_n|^{d-2}} +
\fft{1}{|\bx - \bd_n|^{d-2}}\Big]= 1+ \sum_{n\neq 0} \fft{w_{n}}{r_n^{d-2}}\,. \label{CDWint}
\eea
The total mass of wormhole is given by
\be
\fft{M}{2} = \sum_{n \neq 0} \left[ \fft{d-2}{4(d-1)} \left[\, c_{In} + d_{In}\,\right]  + \gamma\,w_{In} \right]  \,.
\ee

In the limit 
\be
r_n \rightarrow 0, \qquad r_{m} \rightarrow r_{nm} \; \; (m \neq n),
\ee
we get 
\bea
\hat{C}_I & \rightarrow & \left[ \fft{c_{In}}{r_n^{d-2}} + C_{In} \right]\,, \qquad  C_{In} = 1 + \sum_{m \neq n} \fft{c_{Im}}{r_{nm}^{d-2}}\,, \nn \\
\hat{D}_I & \rightarrow & \left[ \fft{d_{In}}{r_n^{d-2}} + D_{In} \right]\,, \qquad  D_{In} = 1 + \sum_{m \neq n} \fft{d_{Im}}{r_{nm}^{d-2}}\,, \nn \\
\hat{W} & \rightarrow & \left[ \fft{w_{n}}{r_n^{d-2}} + W_{n} \right]\,, \qquad \, W_{n} = 1 + \sum_{m \neq n} \fft{w_{m}}{r_{nm}^{d-2}}\,.
\eea
and metric takes the form 
\be
ds^2 \rightarrow \left( \fft{\alpha_n}{r_n^{d-2}} \right)^\ft{4}{d-2} \prod_{I=1}^q \left[ \left( 1 + C_{In} \fft{r_n^{d-2}}{c_{In}} \right) \left( 1 + D_{In} \fft{r_n^{d-2}}{d_{In}} \right) \right]^\ft{1}{d-1} \left[ 1 + W_n \fft{r_n^{d-2}}{w_n} \right]^\ft{ 4 \, \gamma}{d-2} (dr_n^2 + r_n^2 \; d\Omega_{d-1}^2) \,,
\ee
where 
\be
\alpha_n = w_n^\gamma \prod_{I=1}^q (c_{In} \, d_{In})^\ft{d-2}{4(d-1)} \,. 
\ee
From \ref{CDWint} we can see that 
\be
c_{In} = c^{d-2} \,\fft{e^{-n\lambda_I}}{(\sinh n \mu_0)^{d-2}}\,, 
\qquad d_{In} =  c^{d-2} \,\fft{e^{n\lambda_I}}{(\sinh n \mu_0)^{d-2}}\,, 
\qquad w_n = \fft{c^{d-2}}{(\sinh n \mu_0)^{d-2}} \,,
\ee
%%%%%
and hence 
%%%%%
\be
(\,c_{In} \, d_{In} \,) = w_n^2 \, \quad \implies  \quad \alpha_n = w_n^\gamma \prod_{I=1}^q (c_{In} \, d_{In})^\ft{d-2}{4(d-1)} = w_n \,,   
\ee
%%%%%
i.e.
%%%%%
\be
\alpha_n^2 = w_n^2 = c_{In} \, d_{In} \,.
\ee
%%%%%
Now, defining a new radial coordinate 
\be
r_n'^{d-2} = \fft{\alpha^2}{r_n^{d-2}}\,,
\ee
%%%%%
the line element in the limit $r_n' \rightarrow \infty$ takes the form 
\be
ds^2  \rightarrow  \prod_{I=1}^q \left[ \left( 1 + C_{In} \fft{d_{In}}{r_n'^{d-2}} \right) \left( 1 + D_{In} \fft{c_{In}}{r_n'^{d-2}} \right) \right]^\ft{1}{d-1} \left[ 1 + W_n \fft{w_n}{r_n'^{d-2}} \right]^\ft{ 4 \, \gamma}{d-2}  (dr_i'^2 + r_i'^2 \; d\Omega_{d-1}^2)\,. 
\ee
%%%%%
The mass of the $n^{th}$ image point in this system is 
\be
{\mathfrak m}_n = \fft{d-2}{2(d-1)} \left[\, C_{In}\,d_{In} + 
D_{In}\,c_{In}\,\right]  + 2\gamma\,W_{In}\,w_{In}\,,
\ee
or 
\be
{\mathfrak m}_n = \fft{d-2}{2(d-1)} \left[\, c_{In} + d_{In}\,\right]  + 
2 \gamma\,w_{In} + \fft{d-2}{2(d-1)} \sum_{m\neq n} \left[\, 
\fft{c_{In}\,d_{Im} + d_{In}\,c_{Im}}{r_{mn}^{d-2}} \right] + 
2 \gamma \,  \sum_{m \neq n} \, \fft{w_n \, w_m}{r_{mn}^{d-2}}\,.
\ee
Following the discussion similar to previous subsection, the 
``renormalised mass" of the wormhole mouth is 
\be
m_2 = m_1 = \fft{M}{2} + \fft{d-2}{d-1}\, c^{d-2}\,\sum_{I=0}^{q} \, \sum_{p \ge 1}  (p-1)\fft{\cosh p\lambda_I}{[\sinh p\mu_0]^{d-2}} + 2 \,\gamma\, c^{d-2}\,\sum_{p \ge 1}  \fft{p-1}{[\sinh p\mu_0]^{d-2}}\,.
\ee
The interaction energy is then given by 
\be 
M_{int} = M - m_1 - m_2 = - \fft{4(d-2)}{(d-1)} \, c^{d-2}\, \sum_{I=1}^q \sum_{p\ge1} (p-1) \fft{\sinh^2\ft{p\lambda_I}{2}}{(\sinh(p\mu_0)^{d-2}} -
4c^{d-2}\, \sum_{p\ge1} \fft{p-1}{(\sinh p \mu_0)^{d-2}}\,.
\ee

  In a similar fashion, in the case of two Maxwell fields and a single
dilaton, described in section \ref{E2MDsec}, the interaction energy is  
given by
\be
M_{int}= - \fft{2(d-2)}{(d-10}\, c^{d-2}\, \sum_{p\ge 1} 
\fft{p-1}{(\sinh p\mu_0)^{d-2}}\,
\Big[ N_1\, \cosh p\lambda_1 + N_2\, \cosh p\lambda_2\Big]\,.
\ee

For the Einstein-Maxwell-Dilaton system discussed in section \ref{EMDsec}, 
the interaction energy is 
\be
M_{int}= - \fft{16(d-2)\, c^{d-2}}{(d-1)\Delta}\, \sum_{n\ge1} (n-1)
\fft{\sinh^2\ft{n\lambda}{2}}{(\sinh(n\mu_0)^{d-2}} -
4c^{d-2}\, \sum_{n\ge1} \fft{n-1}{(\sinh n\mu_0)^{d-2}}\,.
\ee
For the Einstein-Maxwell case, described in section
\ref{EMsec}, the interaction energy is
\be
M_{int} = - 4\, c^{d-2}\, \sum_{n\ge 1} (n-1) 
 \,\fft{\cosh n\lambda}{(\sinh n\mu_0)^{d-2}}\,.
\ee

\section{Conclusions}

  The geometrodynamical approach to studying solutions of the Einstein
equations and the coupled Einstein-Maxwell equations was pioneered by
Wheeler, Misner and others in late 1950s and early 1960s.  The idea was
to look at the initial-value constraints in a Hamiltonian formulation
of the equations of motion, extracting as much information as possible
about the properties of the (in general time-dependent) solutions that
would evolve from the initial data.  For simplicity, the initial
data were typically taken to be time independent, corresponding to an initial
slice at a moment of time-reflection symmetry in the subsequent evolution.  
One can calculate some general features of the solutions that will
evolve from the initial data, even though in practice the explicit solution
of the evolution equations is beyond reach.
The early work on geometrodynamics was all focused on the case of
four-dimensional spacetimes.

   More recently, wider classes of four-dimensional theories were considered,
in which additional matter fields of the kind occurring in supergravity
theories were included \cite{ortin,cvegibpop}.   
In this paper, we have presented results for time-symmetric initial
value data satisfying the constraint equations in higher-dimensional
theories of gravity coupled to scalar and Maxwell fields.  These theories
encompass particular cases that correspond to higher-dimensional theories
of supergravity, and thus they also have relevance for the low-energy limits
of string theories or M-theory.  We considered initial data both 
for multiple black hole
evolutions and also for wormhole spacetimes.  In the case of wormhole 
spacetimes, we studied some of the properties of the solutions
in detail, including the masses and charges associated with the individual
wormhole throats in a multi-wormhole spacetime, and the interaction energies
between the throats.

  Our focus in this paper has been the construction of consistent time-symmetric
initial data for multiple black holes or wormholes in higher-dimensional 
theories such as those that arise in supergravities or in string theory and
M-theory.  In general, one does not expect to be able to solve the
evolution equations for the initial-data sets explicitly, but it could
nonetheless be of interest to try investigate further some of the features
that might be expected to arise in such solutions.

  A further point is that the solutions to the initial-value constraints 
that we considered all 
made use of an ansatz introduced first by Lichnerowicz in the case of
four-dimensional spacetimes, in which the spatial metric on the 
initial surface is taken to be a conformal factor times a fixed 
fiducial metric of high symmetry, such as the Euclidean metric, or the
metric on $S^3$ or $S^1\times S^2$.  When one considers higher spacetime
dimensions, such a conformal factor parameterises a smaller fraction
of the total space of possible spatial geometries.  It might therefore be
interesting to explore more general ans\"atze for parameterising the
spatial metrics on the initial surface.

\section*{Acknowledgments}

This work was supported in part by DOE grant DE-FG02-13ER42020.

\end{document}